\documentclass[useAMS,usenatbib,twocolumn]{mnras}
\usepackage{natbib}
\usepackage{graphicx}
\usepackage{xcolor}
\usepackage{times}
\usepackage{rotate}
\usepackage{color}
\usepackage{amsmath,bm}
\usepackage{url} 
\usepackage{mathtools}
\usepackage{amsmath}
\usepackage{graphicx}
\usepackage{subfig}
\usepackage{lineno}
\usepackage{comment}
\usepackage{hyperref}
\usepackage{enumitem}
\usepackage{placeins}

\usepackage{soul}

\newcommand{\simgt}{\lower.5ex\hbox{$\; \buildrel > \over \sim \;$}}
\newcommand{\simlt}{\lower.5ex\hbox{$\; \buildrel < \over \sim \;$}}

\newcommand{\rmd}{{\rm d}}
\newcommand{\Msol}{\mbox{$M_{\odot}$}}

\newcommand{\amon}[1]{\textcolor{red}{#1}}

\title[Consistent lensing and clustering in a low-$S_8$ Universe]{Consistent lensing and clustering in a low-$S_8$ Universe \\ with BOSS, DES Year 3,  HSC Year 1 and KiDS-1000}
\author[ A.~Amon, N.~C.~Robertson et al.]{
\parbox{\textwidth}{
\Large{
A.~Amon,$^{1, 2}$\thanks{E-mail: alexandra.amon@ast.cam.ac.uk}
N. C.~Robertson,$^{1,2}$\thanks{E-mail: ncr@ast.cam.ac.uk}
H.~Miyatake,$^{3,4}$
C.~Heymans,$^{5,6}$
M.~White,$^{7,8}$
J.~DeRose,$^{8}$ 
S.~Yuan,$^{9}$
R.~H.~Wechsler,$^{9,10,11}$ 
T.~N.~Varga,$^{12,13}$
S.~Bocquet,$^{14}$
% ...
}
\vspace{0.1cm}
\parbox{\textwidth}{ \small
\textit{The authors' affiliations are shown at the end of this paper. }}
\vspace{-0.6cm}
}}

\date{Accepted XXX. Received YYY; in original form ZZZ}
\pubyear{2022}

\begin{document}
\label{firstpage}
\pagerange{\pageref{firstpage}--\pageref{lastpage}}
\maketitle

\begin{abstract} 
We evaluate the consistency between lensing and clustering based on measurements from BOSS combined with galaxy--galaxy lensing from DES-Y3, HSC-Y1, KiDS-1000. We find good agreement between these lensing datasets. We model the observations using the Dark Emulator and fit the data at two fixed cosmologies: Planck ($S_8=0.83$), and a $\textit{Lensing}$ cosmology ($S_8=0.76$).  For a joint analysis limited to large scales, we find that both cosmologies provide an acceptable fit to the data. Full utilisation of the higher signal--to--noise small-scale measurements is hindered by uncertainty in the impact of baryon feedback and assembly bias, which we account for with a reasoned theoretical error budget. We incorporate a systematic inconsistency parameter for each redshift bin, $A$, that decouples the lensing and clustering.  With a wide range of scales, we find different results for the consistency between the two cosmologies. Limiting the analysis to the bins for which the impact of the lens sample selection is expected to be minimal, for the $\textit{Lensing}$ cosmology, the measurements are consistent with $A$=1; $A=0.91\pm0.04$ ($A=0.97\pm0.06$) using DES+KiDS (HSC). For the Planck case, we find a discrepancy: $A=0.79\pm0.03$ ($A=0.84\pm0.05$) using DES+KiDS (HSC). We demonstrate that a kSZ-based estimate for baryonic effects alleviates some of the discrepancy in the Planck cosmology. This analysis demonstrates the statistical power of small-scale measurements, but caution is still warranted given modelling uncertainties and foreground sample selection effects.
\end{abstract}

\section{Introduction}
The cold dark matter (CDM) model makes precise predictions about the large-scale structure properties of the Universe. In our modern understanding of galaxy formation, every galaxy forms within a dark matter halo. 
The formation and growth of galaxies over time is connected to the growth of the haloes in which they form.
Therefore, an understanding of the statistical relationship between galaxies and haloes, the galaxy--halo connection, is essential in forming a comprehensive interpretation of the observed Universe \citep[for a review, see][]{wechsler2018}. The advent of large galaxy surveys provides a new window into both cosmological and galaxy formation studies \citep{Weinberg2012}, and these two are intertwined. Thus, in order to glean maximal cosmological information from these surveys, it is critical to correctly model the connection between galaxies and their underlying dark matter haloes.

Weak gravitational lensing measures the deflection of light from distant \textit{source} galaxies due to the gravitational potential of matter along the line of sight. More specifically, the weak lensing signal of background galaxies by the intervening matter surrounding foreground \text{lens} galaxies is known as ‘\textit{galaxy--galaxy lensing}’, hereafter, GGL. This signal is therefore correlated with the properties of the `lens' sample and the underlying dark matter large-scale structure it traces. Since its first detection \citep{Brainerd:1996}, this measurement has matured in methodology and signal-to-noise, owing to the wealth of data in the last decade. In particular, the Baryon Oscillation Spectroscopic Survey \citep[BOSS;][]{allam20} and on-going lensing surveys: the Dark Energy Survey\footnote{https://www.darkenergysurvey.org/} \citep[DES;][]{des-overview}, the ESO Kilo-Degree Survey\footnote{http://kids.strw.leidenuniv.nl/} (KiDS; \citealt{Kuijken:2015}), the Hyper Suprime-Cam Subaru Strategic Program\footnote{https://hsc.mtk.nao.ac.jp/ssp/} (HSC; \citealt{Aihara:2018}), have made strides in getting a handle on data calibration, systematics control, and analysis methodology since the first lensing surveys.  

A clustering analysis of galaxies and their redshift-space distortions infers masses indirectly from a combination of density and velocity fields, as well as the constraints on the abundance of haloes in a given cosmological model. Complementary to this, galaxy--galaxy lensing measures the mass of the dark matter haloes around galaxies, tying the galaxies to the underlying dark matter distribution. As such, joint analyses of these two probes have been used to constrain cosmological parameters in the late time Universe \citep*[e.g.][]{seljak2004,Cacciato2009,Cacciato:2013,Mandelbaum_2013,Coupon:2015, More:2015,Kwan_2016,Dvornik2018,Singh:2020} and understand the galaxy--halo connection \citep[e.g.][]{Mandelbaum:2006a,Cacciato2009,Baldauf_2010,Leauthaud:2012,Cacciato:2013,van_den_Bosch_2013, Zu:2015}. With the onset of surveys like the Dark Energy Spectroscopic Instrument\footnote{https://www.desi.lbl.gov} \citep[DESI;][]{DESI2013} and Prime Focus Spectrograph\footnote{https://pfs.ipmu.jp} \citep[PFS;][]{PFS}, in tandem with Vera C. Rubin Observatory’s Legacy Survey of Space and Time\footnote{https://www.lsst.org} (LSST), the ESA’s Euclid mission\footnote{https://www.euclid- ec.org}, and the Roman Space Telescope\footnote{https://roman.gsfc.nasa.gov}, joint analyses with GGL are poised to play an important role in cosmological and galaxy formation studies in the coming decade. These analyses have wide-reaching potential to pin down theoretical systematic uncertainties
and to have further constraining power when combined with analyses with the cosmic shear two-point correlation  \citep{Heymans2021, y3-3x2ptkp}.

There has been much discussion in the literature about an intriguing tension between the measurements of the parameter $S_8\equiv \sigma_8 (\Omega_{\rm m}/0.3)^{0.5}$, which corresponds to $\sigma_8$, the linear-theory standard deviation of matter density fluctuations in spheres of radius 8$h^{-1}$Mpc, scaled by the square root of the matter density parameter, $\Omega_{\rm m}$, at low and high redshift. This quantity is persistently measured to be nearly $10\%$ lower in low-redshift data than that from primary anisotropies Cosmic Microwave Background (CMB) \citet{Planck2018} data, derived as
\begin{equation}
S_8 = 0.834^{+0.016}_{-0.016} \quad Planck \,. 
\end{equation}
Typically, the low redshift analyses either limit to large scales only, or conservatively account for theoretical uncertainties at small scales, which are difficult to model. The effect has been apparent when considering cosmic shear only, with measurements from 
the Canada-France-Hawaii Telescope Lensing Survey \citep[CFHTLenS;][]{Heymans:2013}, and in the most recent constraints from the cosmic shear measurement of HSC \citep{hikage19}, KiDS-1000 \citep{asgari20} and DES Year 3 \citep*[Y3;][]{y3-cosmicshear1,y3-cosmicshear2} which found:
\begin{equation}
\begin{aligned}
S_8 & = 0.800^{+0.029}_{-0.028} \quad \mathrm{HSC \,\, Y1}\\
S_8 & = 0.759^{+0.024}_{-0.021} \quad \mathrm{KiDS-1000}\\ 
S_8 & = 0.772^{+0.018}_{-0.017} \quad \mathrm{DES \,\, Y3} \, .
\end{aligned}
\end{equation}
This `low-$S_8$' has also been evident in joint GGL and clustering work (\citealt{Mandelbaum_2013,Cacciato:2013,HSC2x2}; for \citealt{More:2015}, this was not the case), including those where small-scale systematics are removed by using modified statistics  \citep[e.g.][]{Reyes:2010, Blake:2016, Amon2018, Wibking_2019,Singh:2020, Blake:2020} and in 
findings from a joint analysis of cosmic shear, GGL and clustering measurements, \citep{Joudaki_2017, van_Uitert_2018,DESY13x2,Heymans2021, y3-3x2ptkp}. In addition to constraints from galaxy weak lensing, tension with the primary CMB constraints has recently been found through analyses using \textit{Planck} CMB lensing cross-correlation measurements, using photometry from unWISE \citep{unwiseAK} and the DESI Legacy Survey \citep{Hang2021,Kitanidis2021,white2021cosmological}. At the same time, there are hints of a lower than expected amplitude of low-redshift fluctuations in analyses of the redshift space galaxy power spectrum using BOSS \citep{Damico2020,Ivanov2020, troester20, kobayashi21, Chen22, Philcox22}.

Recent analyses that compare the observed small-scale GGL signal to the prediction based on model fits to projected clustering measurements, when adopting a \textit{Planck} cosmology, have found this consistency test to fail, often referred to as `\textit{Lensing is Low}'. 
\citet[][]{Leauthaud:2017aa} found the \textit{Planck}-prediction from the clustering is larger than the observed CFHTLenS and Canada
France Hawaii Telescope Stripe 82 Survey (CS82) lensing signal around BOSS CMASS Luminous Red Galaxy (LRG) sample by up to 40\%. \citet{Lange2019} confirmed this with the CFHTLenS data and extended the finding to include the BOSS LOWZ LRG sample. Similiarly, the effect was found to be present with SDSS lensing data and LOWZ and shown to be relatively independent of galaxy halo mass \citep{Wibking_2019,Lange2021}. There have been attempts to resolve this discrepancy with alternative small-scale bias modelling \citep{Yuan:2020, Yuan:2021b}. One avenue that has been explored is whether conducting the analysis with lower values for $\Omega_{\rm m}$ and $\sigma_8$ would resolve this discrepancy.  \citet{Leauthaud:2017aa} have shown that using cosmological parameters that are 2--3$\sigma$ lower than \citet{Planck-Collaboration:2015aa} would bring the GGL and clustering predictions into agreement, but also highlight the importance of baryonic effects and assembly bias on scales below a few Mpc,  which incur large modeling uncertainties that are unaccounted for. \citet{Lange2021} used SDSS lensing extending to $>50h^{-1}$Mpc to show the amplitude offset between lensing and clustering to be scale-independent, and concluded that neither a `Lensing cosmology' nor baryonic effects and assembly bias can fully explain the data on both small and large scales.

The reliability of cosmological conclusions based on GGL and clustering measurements at small scales ($<5h^{-1}$Mpc) has been limited by the challenges faced when modelling these measurements.
The most popular model of the galaxy--halo connection used in cosmological studies is the
Halo Occupation Distribution model \citep[HOD; e.g.][]{Peacock:2000,Seljak:2000, Berlind:2002}, which, when combined with the halo model, describes the non-linear matter distribution \citep{Seljak:2000, Cooray:2002}. These models have been used as physically informative descriptions of galaxy bias \citep[see][for example]{Desjacques} that assume all galaxies inhabit dark matter haloes in a manner that depends only on a few specific halo properties, even when baryonic impact is considered \citep[e.g.][]{Acuto2021}.

Typically, the HOD model used is simple; for example, it assumes that galaxy occupation is determined solely by halo mass. However, several factors play a role at smaller scales and the combination of these effects must be accounted for. First, the galaxy clustering observable is the true cosmological signal modulated by an uncertain \textit{galaxy bias} function that maps how galaxies trace the underlying total matter distribution; this can be non-linear, non-local, and redshift-dependent. Furthermore, we need to consider galaxy \textit{assembly bias}, the effect that the clustering amplitudes of dark matter haloes depend on halo properties besides mass, as well as  \textit{baryonic effects} on the matter distribution on small scales \citep{Gao:2005, Wechsler:2006}. In addition, even at larger scales, hydrodynamical simulations have recently been used to test simple HOD models and have found the need for more sophisticated HOD models \citep{Hadzhiyska:2021}. One hurdle is that cosmological clustering needs to be accurately distinguished from artificial clustering in the galaxy sample, arising from potentially uncharacterised inhomogeneities in the target selection \citep[e.g.][]{Ross:2012}. In addition, many of the galaxy samples available have complex colour selections, which may require very flexible HOD forms to model. While each of these systematic effects has failed to resolve the reported 20-40\% discrepancy between the lensing and clustering independently \citep{Leauthaud:2017aa, Lange2019, Lange2021, amodeo21,Yuan:2021b}, they have not been considered in combination. 

In this work, we assess the consistency of lensing and clustering, previously studied with data from CFHTLenS, CS82 and SDSS lensing surveys, now using the state-of-the-art DES Y3, KiDS-1000 and HSC lensing data. These new shear data benefit from a significant development in data calibration techniques and include rigorous estimates of the systematic uncertainty associated with shear and redshift estimates, which we account for. We enhance the investigation by using an emulator-based approach for the halo modelling, the {\sc Dark Emulator} \citep{DQ}. This has been previously assessed to be more accurate than analytical models since the emulating process naturally takes into account effects such as nonlinear clustering, nonlinear halo bias, and halo exclusion. The halo model based on {\sc Dark Emulator} \citep{DQHM} enables flexibility to account for complexities related to the small-scale distribution of galaxies, such as incompleteness and miscentering. Furthermore, with the enhanced survey volume afforded by these data, we consider four lens redshift bins and extend measurements to large scales, allowing us to separately fit and assess the scale-dependence of the consistency. This is important to delineate between the more robust linear scales, and the modelling error associated with small-scale signals. 

Alongside the comparison to clustering-based predictions, this work builds upon `\textit{Lensing without borders}', an inter-survey collaboration \citep*{LWB}. This effort exploits the on-sky overlap of existing lensing surveys with BOSS to perform empirically motivated tests for the consistency of lensing surveys using GGL, based upon the framework in \citet{Amonir}. Here, we present new measurements from DES-Y3 and KiDS-1000 and assess their consistency, as well as with measurements using HSC,  that were published in \citet*{LWB}.

The structure of this paper is as follows. Section~\ref{sec:data} summarises the data used in this analysis: spectroscopy from BOSS and lensing photometry from DES, KiDS and HSC. In Section~\ref{sec:th} we briefly review the cosmological interpretation for the projected galaxy clustering, $w_{\rm p}$, and GGL observables, $\Delta\Sigma$, and the connection to the halo model. In Section~\ref{sec:uncertainties} we investigate the systematic theoretical errors that impact the small-scale modelling. Section~\ref{sec:measurements} presents the estimators and measurement methodology for $w_{\rm p}$ and $\Delta\Sigma$, including the combined DES+KiDS result. Section~\ref{sec:joint} assesses a joint analysis of the clustering and lensing results, considering only the easier-to-model larger-scales and discusses the implications of our findings, particularly in the cosmological context. Section~\ref{sec:results} considers the consistency of clustering and lensing measurements, investigating the scale-dependence and the $S_8$ tension, and small-scale galaxy--halo connection effects. Finally, Section~\ref{sec:outlook} explores how these measurements might be combined with other data to learn about the galaxy--halo connection. In the appendices, A: we revisit Lensing without borders I, and assess the consistency of KiDS, DES and HSC lensing measurements, B: we describe the magnification corrections to the lensing measurements, C: we assess variations to the HOD modelling used in this work and D: we show the joint fits to the data including an additional parameter that captures any inconsistency.

\section{Data}\label{sec:data}

\begin{figure*}
\centering
\begin{tabular}{@{}c@{}}
    \includegraphics[width=16cm]{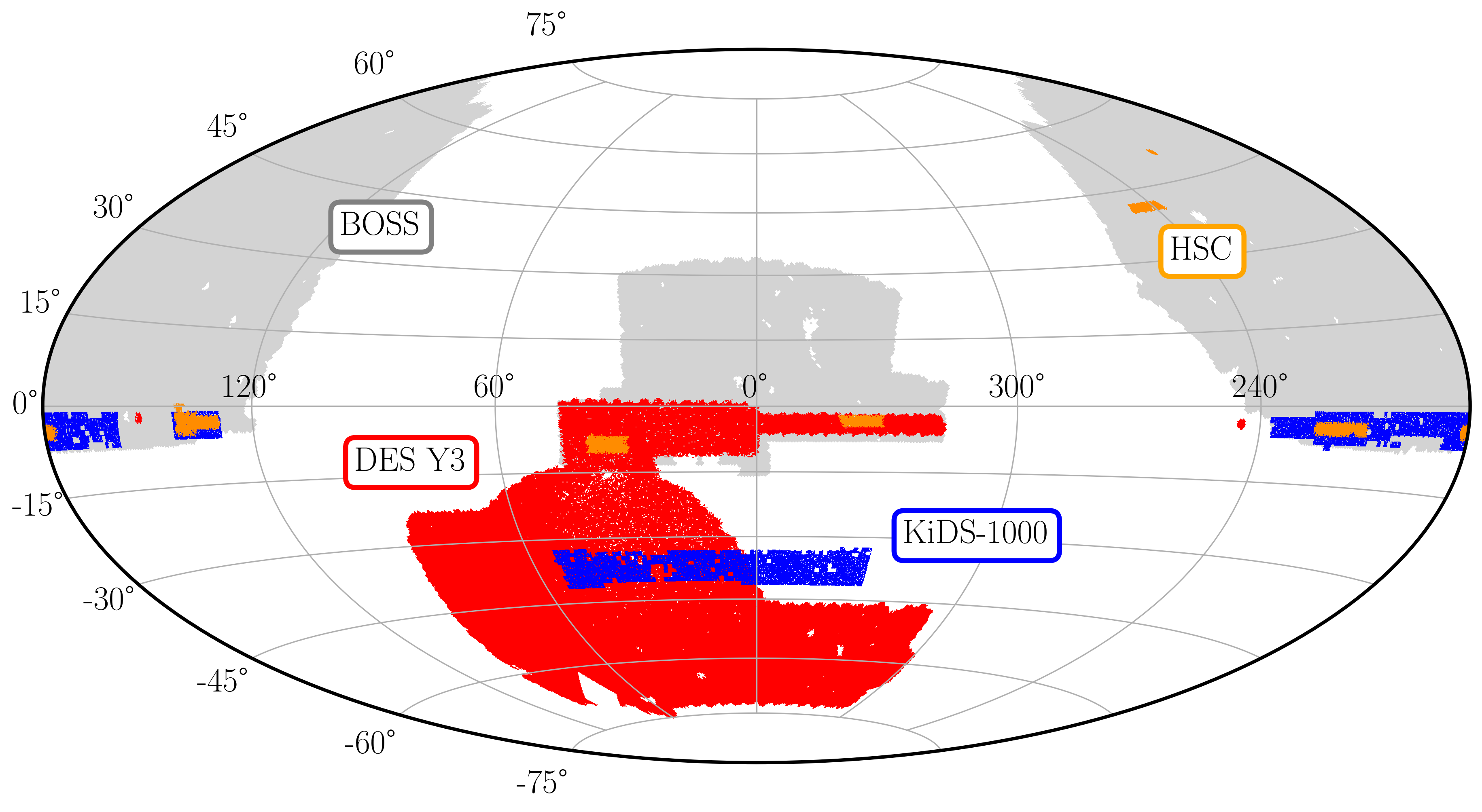}
 \end{tabular}
\caption{The on-sky footprints of the weak lensing surveys considered here: KiDS-1000 (blue), DES Year 3 (red), and HSC Year 1 (yellow), and their overlap with the BOSS redshift survey (grey), which is estimated for each survey as 409, 771 and 137 deg$^2$, respectively.}
\label{fig:map}
\end{figure*}

In this work, the projected galaxy clustering is computed with a foreground spectroscopic `lens' sample, from the SDSS BOSS data. The galaxy--galaxy lensing is measured by stacking
shapes of background galaxies around the foreground lenses, in the region where imaging lensing data overlap on the sky with BOSS. The imaging for the background galaxies is from DES, HSC and KiDS.   The footprints of KiDS-1000, DES-Y3 and HSC-Y1 are illustrated in Figure~\ref{fig:map}, over-plotted on the footprint of the BOSS survey. Their properties are summarised in Table~\ref{wlsurveytable} and their redshift distributions shown in Figure~\ref{fig:nzs}. This section provides brief descriptions on the various data used in this paper, but readers are referred to the original survey papers for more details.

\subsection{BOSS}\label{boss}
BOSS is a spectroscopic survey of 1.5 million galaxies over 10,000 deg$^2$ \citep{Dawson:2013} that was conducted as part of the SDSS-III programme
\citep{Eisenstein:2011} on the Sloan Foundation Telescope (2.5 metre aperture) at Apache Point Observatory \citep{Gunn:1998, Gunn:2006}. BOSS galaxies were selected from Data Release 8 \citep[DR8,][]{Aihara:2011} {\it ugriz} imaging \citep{Fukugita:1996} using a series of colour--magnitude cuts.  
 
BOSS targeted two primary galaxy samples, both of which are used here: the LOWZ sample at $0.15<z<0.43$ and the CMASS sample at $0.43<z<0.7$. 

\begin{figure}
\centering
\includegraphics[width=\columnwidth]{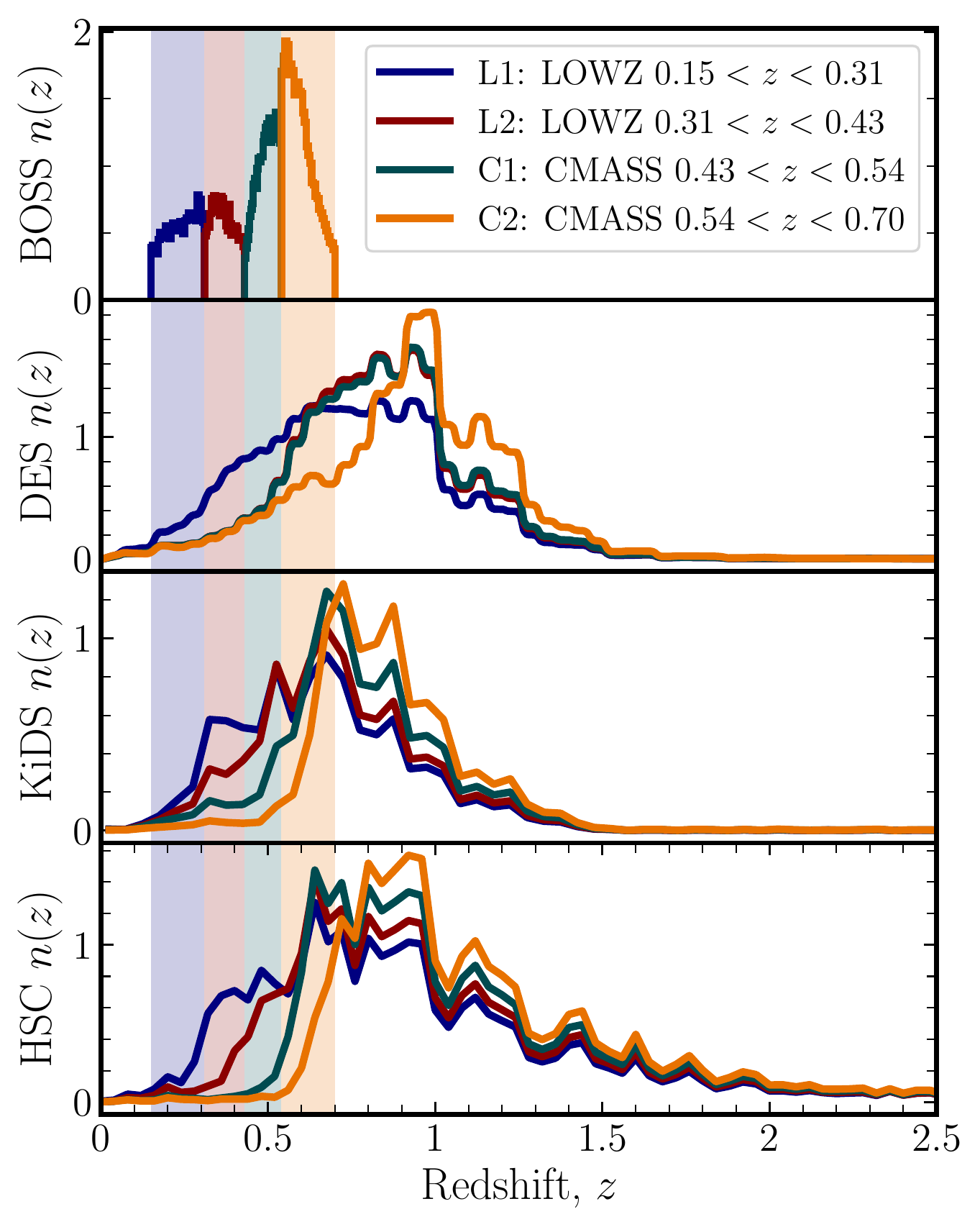}
\caption{The redshift distributions for the BOSS LOWZ and CMASS lenses (upper panel), divided into four distinct samples - L1:$z$=0.15-0.31, L2:$z$=0.15-0.31, C1:$z$=0.31-0.54, C2:$z$=0.54-0.7 with median redshifts of [0.240, 0.364, 0.496, 0.592]. These are the histograms of the samples' redshifts, scaled by $\times 10^3$. The lower panels show the normalised calibrated $n(z)$ for each of KiDS, DES and HSC sources, with the colours corresponding to the subsample used in the measurements with each lens bin, selected to be sufficiently behind the BOSS lenses. For KiDS and HSC, these selections are made using estimates of per-galaxy photometric redshifts, $z_{\rm KiDS/HSC}$, such that  $z_{\rm KiDS/HSC} > z_{\rm l}+0.1$. For DES, the redshift distributions shown are a weighted combination of the calibrated Y3 fiducial source bins. Only those bins that are sufficiently behind the lens bin, that is, for which the mean redshift of the bin, $z_{\rm DES}$, satisfies $z_{\rm DES}>z_{\rm l,max}+0.2$ are used.
}
\label{fig:nzs}
\end{figure}

Following \textit{Lensing without borders} \citep{LWB}, we use Data Release 12 \citep[DR12;][]{Alam:2015} and the large-scale structure catalogs described in \citet[]{Reid:2016} in this work. We divide each of LOWZ and CMASS data into two distinct lens samples by redshift, with bounds:
\begin{eqnarray*}
    {\rm L1: LOWZ} & z=0.15-0.31  \\
    {\rm L2: LOWZ} & z=0.31-0.43  \\
    {\rm C1: CMASS} & z=0.43-0.54  \\
    {\rm C2: CMASS} & z=0.54-0.70  \,.
\end{eqnarray*} 
The redshift distributions, $n(z)$, for these four samples are shown in the upper panel of Figure~\ref{fig:nzs}. 
We incorporate weights for each BOSS galaxy designed
to minimise the impact of artificial observational effects that can bias estimates of the true galaxy overdensity field. This  ensures that the distribution of the randoms traces the variations in the lens samples. Following \citet{Reid:2016}, for LOWZ, this weight is
$w_{\rm z}=w_{\rm cp}+w_{\rm noz}-1$,
where $w_{\rm cp}$ accounts for galaxies that did not obtain redshifts due to fibre collisions by up-weighting the nearest galaxy from the same target class and $w_{\rm noz}$ is designed to account for galaxies for which the spectroscopic pipeline failed to obtain a redshift and upweights in the same manner as $w_{\rm cp}$.
For CMASS, we incorporate additional weights to account for variations in the stellar density and seeing, employed  as, $w_{\rm systot}w_{\rm z}$, where $w_{\rm systot}=w_{\rm star}w_{\rm see}$, such that $w_{\rm star}$ accounts for variations in the CMASS number density with stellar density and $w_{\rm see}$ corrects for variations in the CMASS sample in the seeing \citep{Ross:2012}. 
Note that for the CMASS measurements of projected clustering, we use only the $w_{\rm systot}$ weight and a correction that accounts for both fibre collision and redshift failure \citep{Guo:2012aa,Guo2018} is employed. Similarly, the LOWZ clustering measurements use no weights.

\begin{figure}
\centering
\includegraphics[width=\columnwidth]{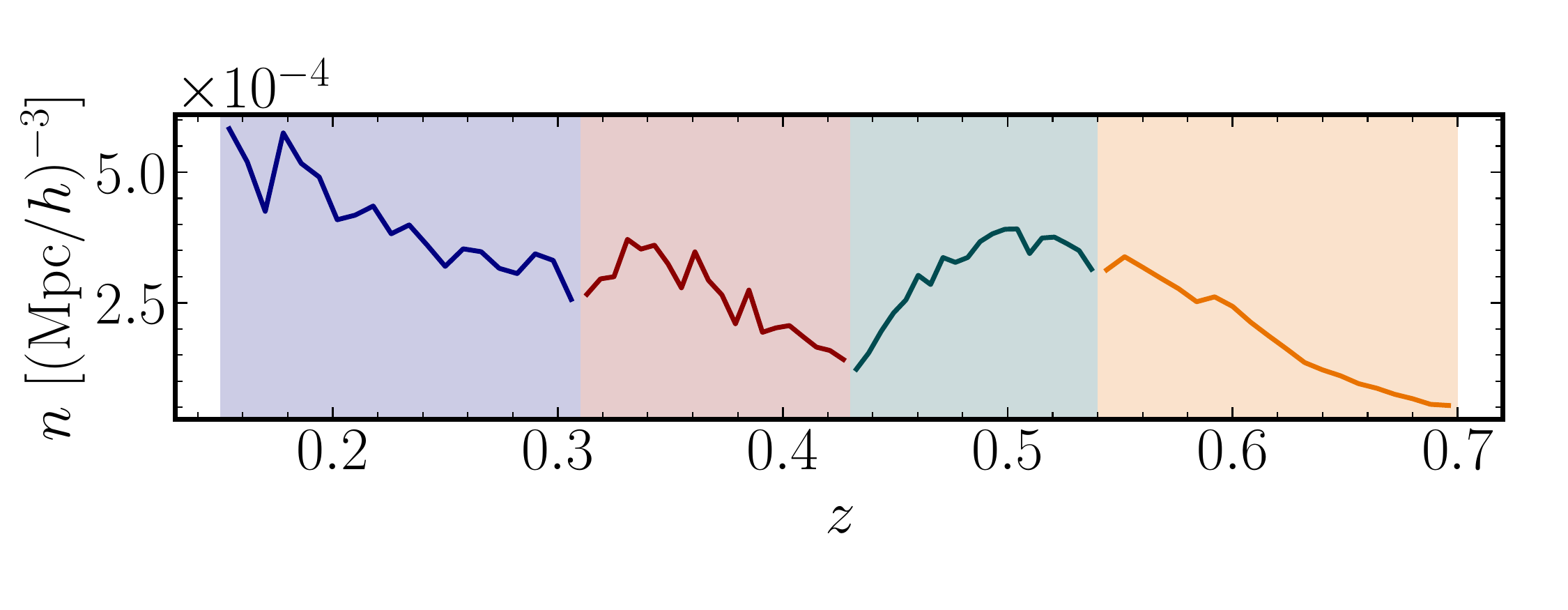}
\caption{The LOWZ and CMASS galaxy samples in the plane of redshift, divided into four distinct samples: L1, L2, C1, C2  by redshift. The solid lines show the redshift dependence of the comoving number density of galaxies in each of our galaxy samples, obtained assuming the \textit{Planck} cosmology.
The number density depends on redshift within each sample, with the C1 sample showing a somewhat stronger redshift dependence, uniquely with a number density that increases with redshift, as a consequence of that bin's particular selection effects.}
\label{fig:nz_pervol}
\end{figure}

The BOSS LOWZ and CMASS galaxy samples are colour selected: The LOWZ sample primarily selects red galaxies and the CMASS sample targets galaxies at higher redshifts with a surface density of roughly 120 deg$^{-2}$ \citep[][]{Reid:2016}. Our model described in Section~\ref{sec:incom} can account for stellar mass incompleteness if it is solely dependent on the halo mass
\citep{More:2015}. However, it assumes that any other sample selection criteria do not correlate with the large-scale environment at fixed halo mass. Thus our model may not be an entirely accurate description in the presence of color-based galaxy assembly bias coupled with color-based selection effects.
Towards the lower end of the redshift range, the colour cuts remove star-forming galaxies and have a bigger impact on the ${n}(z)$ of the sample, whereas the higher-redshift samples, C2 and L2, are closer to being flux limited and includes a larger range of galaxy colours at fixed magnitude \citep{Reid:2016}. We show in Figure~\ref{fig:nz_pervol} that this is especially true for the C1 sample, which increases in number density as a function of redshift.  As simple mass dependent HODs are designed for complete galaxy samples, this brings into question whether a simple HOD is sufficient for modeling these lower-redshift L1 and C1 samples. Although we divide the BOSS galaxies into four redshift bins, the target selection is such that the samples are likely still evolving across the redshift range that they span. In this case, a fixed HOD model is a simplification.


\begin{table}
  \centering
  \caption{Properties of weak lensing surveys used in this paper. We quote the survey area [deg$^2$] (after masking), the overlap area with BOSS, the number of lenses in the overlap area for the two LOWZ samples (L1 and L2) and the two CMASS samples (C1 and C2), the median unweighted redshifts of the source distribution, and the effective weighted galaxy number density measured in galaxies per square arcminute (see Equation 1 in \citealt{Heymans:2012}) after photometric redshift quality cuts.}
\begin{tabular}{@{}lcccc}
\hline
& HSC-Y1 & DES-Y3 & KiDS-1000  \\
\hline
Area [deg$^2$]& 137 &  4143  & 777  \\
BOSS overlap area & 137  & 771 & 409  \\
Num. LOWZ (L1/L2) & 3170 / 3251  & 16285 / 17661 & 4701 / 5431 \\
Num. CMASS (C1/C2)  & 8409 / 8943  & 30613 / 34223  & 17501 / 18509  \\
$z_{\rm med}$   & 0.80  & 0.63  & 0.67  \\
$n_{\rm eff}$  & 21.8  &  5.59  & 6.22  \\
\hline
\end{tabular}
\label{wlsurveytable}
\end{table}

\subsection{DES Year 3}\label{sec:datades}

For this analysis, we use Dark Energy Survey (DES) data taken during the survey's first three years of survey operation (Y3), between 2013 and 2016 \citep{y3-gold}, from the  4-meter Blanco Telescope and using the Dark Energy Camera \citep{Flaugher2015}. The DES Y3 footprint covers 4143 deg$^2$ in five broadband filters ($grizY$). The number density of the DES Y3 data is 5.59 arcmin$^{-2}$, as summarised in Table~\ref{wlsurveytable}, and the dataset has 771 deg$^2$ sky area in common with BOSS.

The shape catalog is created with \textsc{metacalibration} \citep{Sheldon2017, Huff2017} to give over 100 million galaxies that have passed a raft of validation tests \citep{y3-shapecatalog}. The source sample has been divided into four redshift bins and the redshift distributions and associated uncertainty are calibrated primarily using a machine-learning technique, Self Organizing Maps \citep[SOM; ][]{buchs19}, which exploits DES Deep Field data with near-infrared overlap \citep{y3-deepfields}. Remaining biases in the shape measurement and redshift distributions, primarily due to blending,  are calibrated using image simulations, and the associated corrections for each redshift bin are reported in \citet{y3-imagesims}. The outcome of these two methods combined is a set of realisations of the source redshift distributions, for the four redshift bins, which span the uncertainty in the calibration. These include the contribution to the uncertainty arising due to the redshift-mixing impact of blending. 
For the measurements with each BOSS redshift bin, we use only the DES redshift bins that are sufficiently behind the lens sample along the line-of-sight, defined such that the mean redshift of the DES bin, $z_{\rm DES}$, satisfies $z_{\rm DES}>z_{\rm l,max}+0.2$. For lens bins {L1, L2, C1, C2}, that corresponds to using the DES Y3 tomographic bins {[2,3,4],[3,4],[3,4],[4]}, respectively, which are combined using an inverse-variance weighted average.

\subsection{KiDS-1000}\label{sec:datakids}

The Kilo-Degree Survey is an optical wide-field survey using the OmegaCam camera mounted on the VLT Survey Telescope located at the Paranal Observatory. Observations are made in four bands ($ugri$); the VISTA Kilo-degree Infrared Galaxy survey (VIKING) has, by design, observed the same area of sky in an additional five bands ($ZYJHK_s$; \citealt{Edge2013}), making KiDS a deep and wide nine-band imaging data set \citep{wright20}. In this work we use the fourth KiDS data release \citep[][hereafter KiDS-1000]{kuijken19}, which consists of 1006 deg$^2$ of galaxy lensing data. The number density of of the KiDS-1000 sample is 6.22 arcmin$^{-2}$, over a net masked area of 777 deg$^2$, of which $409$ deg$^2$ covers the BOSS footprint. 

The source galaxies are selected using the best-fit photometric redshift, $z_{\rm B}$, determined from the nine-band imaging for each source using the Bayesian code {\tt BPZ} \citep{Benitez:2000}. In this analysis, the redshift distributions are calibrated following the same approach adopted in the KiDS-1000 cosmology analyses \citep{asgari20,Heymans2021}, which uses a SOM \citep{SOM} to define a source sample of only those galaxies whose redshift is accurately calibrated using spectroscopy \citep{Hildebrandt2021}. Although using the SOM method leads to a reduction in the galaxy number density and therefore a small increase in the statistical error over the same area, here we prioritise minimising the systematic error associated with a photometric redshift distribution. 
Source galaxies are selected to be behind the lens, such that  $z_{\rm B} > z_{\rm l}+0.1$, following tests in \citet{Amonir}, designed to reduce the boost correction. The redshift distribution of the source galaxies behind each lens is then re-estimated and calibrated following the procedure described in Section~\ref{sec:datakids}. 
The shape measurements are computed with the {\it lens}fit pipeline \citep{Miller2013}, calibrated on simulations presented in \citet{kannawadi19, giblin20}.

\subsection{HSC}\label{hsc}
The Hyper Suprime-Cam Subaru Strategic Program aims to cover 1,400~deg$^2$ of the sky in bands $(grizY)$ using the Hyper Suprime-Cam \citep{Miyazaki:2018, Komiyama:2018} on the Subaru 8.2m telescope. The survey design is described in \citet[][]{Aihara:2018}, the analysis pipeline  in \citet[][]{Bosch:2018aa}, and validation tests of the pipeline photometry in \citet[][]{Huang:2018ac}.  The first data release \cite[DR1,][]{Aihara:2018aa}, used in this work, maps an area of 136.9 deg$^2$ (with complete BOSS overlap) split into six fields and has a  mean $i$-band seeing of 0.58$\arcsec$ and a 5$\sigma$ point-source depth of $i\sim26$.

For HSC Y1, the galaxy-shape estimation is derived from $i$-band images using a moments-based method, with details given in \citet[][]{Mandelbaum:2018ab} and shear calibration described in \citet[][]{Mandelbaum:2018aa}. The HSC Y1 shear catalog uses a conservative source galaxy selection including a magnitude cut of $i < 24.5$. The weighted source number density is 21.8 arcmin$^{-2}$. Similar to the case for KiDS, source galaxies are selected to be behind the lens, such that  $z_{\rm B} > z_{\rm l}+0.1$ to reduce the source-lens overlap. Photometric redshifts have been computed with the \texttt{frankenz} hybrid method described in \citet[][]{Speagle:2019aa}, that combines Bayesian inference with machine learning and trains on a catalogue of sources including a combination of spectroscopic, grism, prism, and many-band photometric redshifts. Using the \texttt{best} photo-$z$ value from \citet[][]{Speagle:2019aa}, the source distribution in this paper has a mean redshift of $z_{\rm s}=0.95$ and a median of $z_{\rm s}=0.8$.

\section{Theory} \label{sec:th} 
This section briefly reviews the theoretical expressions for the observables that form the basis of the study. These comprise the auto and cross-correlations between weak gravitational lensing and galaxy overdensity, that is, the galaxy--galaxy lensing and projected clustering signals.

\subsection{Differential surface density}  
Galaxy--galaxy lensing can be expressed in terms of the cross-correlation of a galaxy overdensity, $\delta_{\rm g}$, and the underlying matter density field, $\delta_{\rm m}$: for a fixed redshift, is given by $\xi_{\rm gm}(|\bf{r}|)=\langle \delta_{\rm g}(\bf{x})\delta_{\rm m}(\bf{x}+\bf{r})\rangle_{\bf{x}}$.
The lensing galaxy--matter cross-correlation function, $\xi_{\rm gm}$, can be expressed in terms of its Fourier-transformed counterpart, the galaxy--matter cross power spectrum, $P_{\rm gm}(k)$, as, 
\begin{equation}
\centering   
\xi_{\rm gm}(r, z_{\rm l}) \equiv \int_{0}^{\infty} \frac{k^2 dk}{2\pi} \, P_{\rm gm}(k, z_{\rm l})j_2(kr)\, , 
\end{equation}
where $j_2(kr)$ is the second order spherical Bessel function. In order to measure $\xi_{\rm gm}$, one can first determine the comoving projected surface mass density, $\Sigma_{\rm com}$, around a foreground lens at redshift $z_{\rm l}$,  using a background galaxy at redshift $z_{\rm s}$ and at a comoving projected radial distance from the lens, $R$. This is given as, 
\begin{equation}
\label{eqn:deltasigdef}
\centering
\Sigma_{\rm com}(R, z_{\rm l})= \overline{\rho_{\rm m}} \int_{0}^{\chi(z_{\rm s})}  \ \xi_{\rm gm} \big( \sqrt{R^2+[\chi-\chi(z_{\rm l})]^2} \big) \ \rm{d}\chi \, ,
\end{equation}
  where $\overline{\rho_{\rm m}}$ is the mean matter density of the Universe, $\chi$ is the comoving line-of-sight distance and ${\chi(z_{\rm l})}$, ${\chi(z_{\rm s})}$ are the comoving line-of-sight distances to the lens and source galaxy, respectively. The shear is sensitive to the density contrast; therefore, it is a measure of the excess or differential surface mass density, $\Delta \Sigma_{\rm com}(R)$ \citep{Mandelbaum_2005}. This is defined in terms of $\Sigma_{\rm com}(R)$ as,
\begin{equation}
\Delta \Sigma_{\rm com}(R)=\overline{\Sigma}_{\rm com}(\leq R) - \Sigma_{\rm com}(R) \, ,
\end{equation}
where the average projected mass density within a circle is
\begin{equation}
\label{eqn:apmd}
\overline{\Sigma}_{\rm com}(\leq R)=\frac{2}{R^2} \int^R_0 \Sigma_{\rm com}(R') R' \rm{d}R' \, .
\end{equation}

\subsection{Galaxy Clustering: Projected Correlation Function } 
The 3D galaxy correlation function, $\xi_{\rm gg}(r)$, is related to the auto-power spectrum of the galaxy number density field, $P_{\rm gg}$, by
\begin{equation}
\centering
\xi_{\rm gg}(r, z_{\rm l})=\int^{+\infty}_{0} \frac{k^2 dk}{2\pi} P_{\rm gg}(k, z_{\rm l})j_0(kr) \, .
\end{equation}
Galaxy clustering, which is mostly independent of the redshift-space distortion (RSD) effect due to the peculiar velocities of galaxies measurements, can be analysed in terms of the projected separation of galaxies on the sky. We call the associated two-point function in real space the `projected correlation function', $w_{\rm p}(R)$, and it is formulated from the integral of the 3D galaxy correlation function, $\xi_{\rm gg}(r)$, along the line of sight as,
\begin{equation}
\label{eqn:wp}
\centering
w_{\rm p}(R) \equiv 2 \int^{Z_{\rm max}}_{0}\xi_{\rm gg}(r=\sqrt{R^2+Z^2}) \ dZ \, ,
\end{equation}
where $Z$ is the co-moving separation along the line-of-sight. Note that throughout this paper, $Z_{\rm max}=100h^{-1}$Mpc and the projected correlation function has units of [$h^{-1}$Mpc].

Although the clustering signal becomes less sensitive to RSD after the line-of-sight integration, it is a non-negligible effect. It can be corrected for following \cite{DQHM}, based upon the prescription derived by \cite{van_den_Bosch_2013}, as a multiplicative correction factor $f_{\rm RSD} (R; Z_{\max},\beta)$. Here, $\beta$ is the linear Kaiser factor at redshift $z$, given in terms of the growth rate of structure, $f(z)$ as $\beta(z) \equiv (f(z)/b(z))$, and \cite{DQHM} approximated the bias factor $b$ as an effective bias for a sample of galaxies derived from the {\sc Dark Emulator} output. The RSD correction increases as a function of the projected separation, and the size of correction is typically a few percent at $R\sim 1h^{-1}{\rm Mpc}$ \citep{van_den_Bosch_2013}.

\subsection{The galaxy--halo connection}\label{sec:ghc}

We assume that all galaxies are hosted by dark matter haloes, such that a central galaxy is the large, luminous
galaxy that resides at the centre of the halo and many smaller, less-luminous satellite galaxies exist around it and comprise the non-central part of the galaxy--galaxy lensing signal. The correlation function of matter is given by the contributions from particles in the same halo, the one-halo term, and those in two different haloes, the two-halo term. Furthermore, the occupation of haloes with galaxies is assumed to depend on the halo mass $M_{\rm h}$ only. To connect haloes to galaxies, the halo occupation distribution (HOD) is employed \citep{Jing:1998,Peacock:2000, Scoccimarro:2001}. This defines the total occupation of galaxies, $\langle N | M_{\rm h} \rangle$, in a given halo of mass, $M$, in terms of the mean number of central, $ N_{\rm c}$ and satellite,  $ N_{\rm s}$ galaxies, as
\begin{equation}
	\langle N | M_{\rm h} \rangle = \langle N_{\rm c} | M_{\rm h} \rangle + \langle N_{\rm s} | M_{\rm h} \rangle \, .
\end{equation}
Following \citet{More:2015}, the expected number of centrals in our HOD framework is given as,
\begin{equation}
\langle N_{\rm c} | M_{\rm h} \rangle = f_{\rm inc}(M_{\rm h}) \frac{1}{2} \left[ 1 + \mathrm{erf} \left( \frac{\log M_{\rm h} - \log M_{\rm min}}{\sigma_{\log M_{\rm h}}} \right) \right] \, ,
\end{equation}
where the scatter in the halo mass--galaxy luminosity
relation is parameterised by $\sigma_{\log M_{\rm h}}$ and $M_{\rm min}$ is the mass scale at which the median galaxy luminosity corresponds to the threshold
luminosity. $\mathrm{erf}(x)$ is the error function and $M_{\rm min}$ and $\sigma_{\log M_{\rm h}}$ are free parameters and such that 
$\langle N_{\rm c} | M_{\rm h} \rangle$ goes to zero for low halo masses and increases towards higher halo masses. Note that $f_{\rm inc}$ allows for an overall incompleteness in the target selection of BOSS: for haloes of a fixed mass, not all the central galaxies associated with those haloes will be selected into the lens sample, such that $\langle N_{\rm c} | M_{\rm h} \rangle \xrightarrow{ }f_{\rm inc}$ as $M_{\rm h} \xrightarrow{} \infty$ (see  Section~\ref{sec:incom} and \citealt{More:2015} for details). The fiducial model used in this work has $f_{\rm inc}=1$ (see Appendix~\ref{app:model} for justification). 

We assume the average number of satellites obeys the form
\begin{equation}
	\langle N_{\rm s} | M_{\rm h} \rangle \equiv \langle N_{\rm c} | M_{\rm h} \rangle \lambda_{\rm s}(M) =  \langle N_{\rm c} | M_{\rm h} \rangle
	\left( \frac{M_{\rm h} - \kappa M_{\rm min}}{M_1} \right)^{\alpha} \, .
\end{equation}
In the above parametrisation, we assume that the distribution of central galaxies, $N_{\rm c}$, follows the Bernoulli distribution (i.e., can take only zero or one) with mean $\langle N_{\rm c} | M_{\rm h} \rangle$,  and that the satellite galaxies reside only in haloes that host central galaxies \footnote{In reality some dark matter haloes could host a satellite BOSS galaxy even though their central galaxies are missing in the BOSS galaxy sample due to incompleteness. This possibility is partially addressed by allowing for a non-trivial fraction of mis-centered haloes (see Section~\ref{app:model}).}. We further assume that $\langle N_{\rm s} | M_{\rm h} \rangle$ follows a Poisson distribution with mean $\lambda_{\rm s}(M)$. In addition, $\kappa$, $M_1$ and $\alpha$ are free model parameters. Within this HOD framework, the mean number density of galaxies is obtained via an integral over the halo mass function,
\begin{equation}
	n_{\rm gal} = \int\limits_0^\infty (\langle N_{\rm c} | M_{\rm h} \rangle + \langle N_{\rm s} | M_{\rm h} \rangle) \ n_{\rm h} (M_{\rm h}) \ \rmd  M_{\rm h}\, ,
\end{equation} 
where $n_{\rm h} (M_{\rm h})$ is the halo mass function, which gives the mean number density of halos in the mass range $[M,M+dM]$.

The HOD model used in this analysis has five free parameters: ${M_{\rm min},\sigma_{\log M_{\rm h}}, \kappa, M_1, \alpha}$.
The corresponding prior ranges that are adopted throughout this study are listed in Table~\ref{tab:priors}. 

\subsection{Incomplete galaxy samples}\label{sec:incom}

Spectroscopic lens samples like the BOSS LOWZ and CMASS samples used in this work are comprised of colour-selected galaxies. As such, at fixed stellar mass, CMASS is not a random sample of the overall population in terms of galaxy colour, described as colour incompleteness. On the other hand, stellar mass incompleteness describes that some fraction of even very massive haloes might not host a BOSS-like galaxy at low stellar masses compared to a true stellar mass threshold sample \citep{Leauthaud:2016}.  Both of these effects have implications for such analyses, because the HOD form traditionally used is not designed to capture such complex selections.

One variant of the {\sc Dark Emulator} model explored in Appendix~\ref{app:model} attempts to account for stellar mass incompleteness in the selection of BOSS lens galaxies, via the function $f_{\rm inc}(M)$. 
Following \citet{More:2015}, we assume a log-linear functional form for this as follows,
\begin{equation}
f_{\rm inc}(M_{\rm h}) = {\rm max}[0, {\rm min}[1, 1 + \alpha_{\rm inc}({\rm log} M_{\rm h} - {\rm log} M_{\rm inc})]]\, ,
\end{equation}
which explicitly assumes that BOSS selects a random fraction of the stellar mass threshold galaxies from host haloes at every mass scale, defined in terms of two parameters, $\alpha_{\rm inc}$ and ${\rm log} M_{\rm inc}$. That is, it assumes that with the complex BOSS colour and magnitude cuts, the selection probability for galaxies at a given stellar mass only depends upon the halo mass, and not on the environment or other astrophysical properties. Given the colour incompleteness of the sample is not accounted for, even in the absence of assembly bias and other intricate effects of the galaxy--halo connection, it is likely that a more flexible HOD form is needed to fit such galaxies if we want highly accurate results. 

\subsection{Implementation: The {\sc Dark Emulator}}\label{sec:thcorr}

In this section we describe details of the theoretical template and halo model implementation used to obtain model predictions for the observables, $\Delta\Sigma(R)$ and $w_{\rm p}(R)$, for a given cosmological model within the $\Lambda$CDM framework.
Primarily, the analysis takes an emulator approach, using the {\sc Dark Emulator}\footnote{\url{https://github.com/DarkQuestCosmology/dark_emulator_public}}, based on dark matter-only N-body simulations populated with galaxies \citep{DQ,DQHM}. In Section~\ref{sec:uncertainties}, a comparison to a simulation-based prediction is investigated.

\begin{table}
\centering
\begin{tabular}{cc}
	\hline
	Parameter & Prior \\
	\hline
	 $\log M_{\rm min}$ & $[12.0, 14.5]$\\
	 $\sigma_{\log M_{\rm h}}$ & $[0.01, 1.0]$ \\
	 $\log M_1$ & $[12.0, 16.0]$ \\
	 $\alpha$ & $[0.5, 3.0]$ \\
	 $\kappa$ & $[0.01, 3.0]$ \\
	\hline
\end{tabular}
\caption{The five HOD parameters used in this analysis and their priors, chosen following \citet{HSC2x2}. Priors are uniform where $[a, b]$ indicates the upper and lower bounds. }
\label{tab:priors}
\end{table}

The {\sc Dark Emulator} enables a fast, accurate computation of halo statistics 
(halo mass function, halo auto-correlation function and halo--matter cross-correlation) as a function of halo mass (for $M_{\rm 200m} \simgt 10^{12}h^{-1}M_\odot)$, redshift, separation and cosmological parameters under the flat $w$CDM model. The emulator is based on the Principal Component Analysis and Gaussian Process Regression for the large-dimensional input data vector of an ensemble of N-body simulations. Each of these DarkQuest simulations were constructed from 2048$^3$ particles within a box size of 1 or $2\,h^{-1}$Gpc, for 100 flat $w$CDM cosmological models sampled based on a maximum-distance Sliced Latin Hypercube Design. 

For each simulation realisation, for a given cosmological model, a catalogue of haloes was extracted using {\sc Rockstar} \citep{Behroozi:2013}. This identifies haloes and subhaloes based on the clustering of N-body particles in position and velocity space. The spherical overdensity mass, defined with respect to the halo center, signified by the maximum mass density, $M \equiv M_{\rm 200m} = (4\pi/3)R^3_{\rm 200m} \times (200\overline{\rho}_{\rm m})$, is used to define halo mass, where $R_{\rm 200m}$ is the spherical halo boundary radius within which the mean mass density is $200 \times \overline{\rho}_{\rm m}$. The emulator is built upon  the halo catalogs extracted at multiple redshifts in the range of $z = [0, 1.48]$. In this work, for the bins [L1, L2, C1, C2], we assume fixed redshifts at the median of the samples, [0.240,0.364,0.496,0.592].

The accuracy of the {\sc Dark Emulator} predictions has been validated in \cite{DQ}. The halo mass function has 1-2\% accuracy for halos with $M>10^{12}h^{-1}M_\odot$ \footnote{As can be seen in Figure~\ref{fig:hod}, the halo occupation distribution for all our samples falls well below 1\% at the resolution limit of the {\sc Dark Emulator}}, except for the massive end ($M\simgt10^{14}h^{-1}M_\odot$) in which the Poisson error is significant in the simulations used for both training and validations. For haloes of $10^{13}M_{\odot}$, the typical mass of host haloes of BOSS galaxies \citep{White:2011,Parejko:2013,Saito:2016}, the halo--matter cross-correlation function and the halo auto-correlation function have $\sim$2\% accuracy over the comoving separation $0.1h^{-1}{\rm Mpc} < R < 30h^{-1}{\rm Mpc}$ and a degradation to $\sim$3-4\% accuracy at $R \simlt 1h^{-1}{\rm Mpc}$ due to the halo exclusion effect, respectively.

Since the excess surface density is a non-local observable, i.e., the small-scale information of the halo--matter cross-correlation function affects the excess surface density at large scales, the inaccuracy of the halo--matter cross-correlation at $R<0.1h^{-1}{\rm Mpc}$ due to a finite resolution of $N$-body simulations, which was quantified in \cite{DQ}, may affect the excess surface density at $R\sim 0.1h^{-1}{\rm Mpc}$. We explicitly quantify the effect, mimicking the inaccuracy by modifying the halo--matter cross-correlation function from {\sc Dark Emulator} at $r < 0.1h^{-1}{\rm Mpc}$ by a similar amount shown in Figure~7 in \cite{DQ}, and find that the excess surface density based on the modified halo--matter cross correlation function has only a few percent shift at $R \sim 0.1h^{-1}{\rm Mpc}$. Thus we conclude that the excess surface density of {\sc Dark Emulator} has sufficient accuracy on relevant scales for this work, compared to the statistical uncertainties of our measurements.

\subsection{Modelling the correlation functions} 
As seen in Section 2.1 and 2.2, in order to compute the observables, we need the cross-power spectrum of galaxies and matter $P_{\rm gm}(k; z)$, and the real-space auto-power spectrum of galaxies, $P_{\rm gg}(k; z)$, as a function of the parameters of halo-galaxy connection and the cosmological model. Within the halo model, the galaxy--matter cross-power spectrum is given as
\begin{multline}
    P_{\rm gm}(k;z)=\frac{1}{\overline{n}_{\rm g}} \int {\rm d}M_{\rm h} \frac{{\rm d}n_{\rm h}}{{\rm d}M_{\rm h}} \\
    \Big[\langle N_{\rm c} | M_{\rm h} \rangle + \langle N_{\rm s} | M_{\rm h} \rangle \tilde{u}_{\rm s}(k; M_{\rm h},z)\Big]  
    \times P_{\rm hm}(k; M_{\rm h},z) \, ,
    \label{eqn:pgm}
\end{multline}
where $P_{\rm hm}(k; M_{\rm h},z)$ is the halo--matter cross-power spectrum and $\tilde{u}_{\rm s}(k; M_{\rm h},z)$
is the normalised Fourier transform of the averaged radial profile of satellite galaxies, assumed to be a truncated NFW profile, in host haloes of mass, $M_{\rm h}$
at redshift, $z$. In practice this is evaluated at the mean lens redshift, which is weighted appropriately to match the weighting scheme in the excess surface density estimator. This has been shown to be equivalent to using the full lens redshift distribution \citep{HSC2x2}.
The {\sc Dark Emulator} derives $\xi_{\rm hm}(r;M_{\rm h},z)$, the Fourier transform of $P_{\rm hm}(k; M_{\rm h},z)$, and the halo mass function ${\rm d}n_{\rm h}/{\rm dM_{\rm h}}$ for an input set of parameters (halo mass, separation, and cosmological parameters).
In order to obtain real-space $\xi_{\rm gm}$ or $\Delta\Sigma$ for the assumed model, the publicly available code, {\sc FFTLog}, \citep{Hamilton:2000} is used to perform the Hankel transforms.

To model $w_{\rm p}$, the auto-power spectrum of galaxies in a sample is decomposed into the two contributions, the one- and two-halo terms, and those are given within the halo model framework as
\begin{equation}
    P_{\rm gg}(k;z)=P_{\rm gg}^{\rm 1h}(k;z)+P_{\rm gg}^{\rm 2h}(k;z)
\end{equation}
where
\begin{multline}
    P_{\rm gg}^{\rm 1h}(k;z)=\frac{1}{\overline{n}_{\rm g}^2} \int {\rm d}M_{\rm h} \frac{{\rm d}n_{\rm h}}{{\rm d}M_{\rm h}}\langle N_{\rm c} | M_{\rm h}\rangle \\
    \Big[2\lambda_{\rm s}(M_{\rm h})\tilde{u}_{\rm s}(k; M_{\rm h},z)+\lambda_{\rm s}(M_{\rm h})^2\tilde{u}_{\rm s}(k; M_{\rm h},z)^2\Big]\, ,
    \label{eqn:pgg1h}
\end{multline}
\begin{multline}
    P_{\rm gg}^{\rm 2h}(k;z)=\frac{1}{\overline{n}_{\rm g}^2} \Big[\int {\rm d}M_{\rm h} \frac{{\rm d}n_{\rm h}}{{\rm d}M_{\rm h}}\langle N_{\rm c} | M_{\rm h} \rangle (1+\lambda_{\rm s}(M)\tilde{u}_{\rm s}(k; M_{\rm h},z))\Big] \\ \times  \Big[\int {\rm d}M'_{\rm h} \frac{{\rm d}n_{\rm h}}{{\rm d}M'_{\rm h}}  \langle N_{\rm c} | M_{\rm h}' \rangle (1+\lambda_{\rm s}(M_{\rm h})\tilde{u}_{\rm s}(k; M_{\rm h},z))\Big]\\
    \times P_{\rm hh}(k,M_{\rm h},M'_{\rm h},z) \, ,
    \label{eqn:pgm2h}
\end{multline}
where $P_{\rm hh}(k,M_{\rm h},M'_{\rm h},z)$
is the power spectrum between two halo samples with masses $M$ and $M'$. The {\sc Dark Emulator} outputs the real-space correlation function of haloes and
$\xi_{\rm hh}(k,M_{\rm h},M'_{\rm h},z)$, the Fourier transform of $P_{\rm hh}$. The details of the halo model prescription implemented in the {\sc Dark Emulator} can be found in \cite{DQ} and \cite{DQHM}.

\begin{figure*}
\centering
\includegraphics[width=\textwidth]{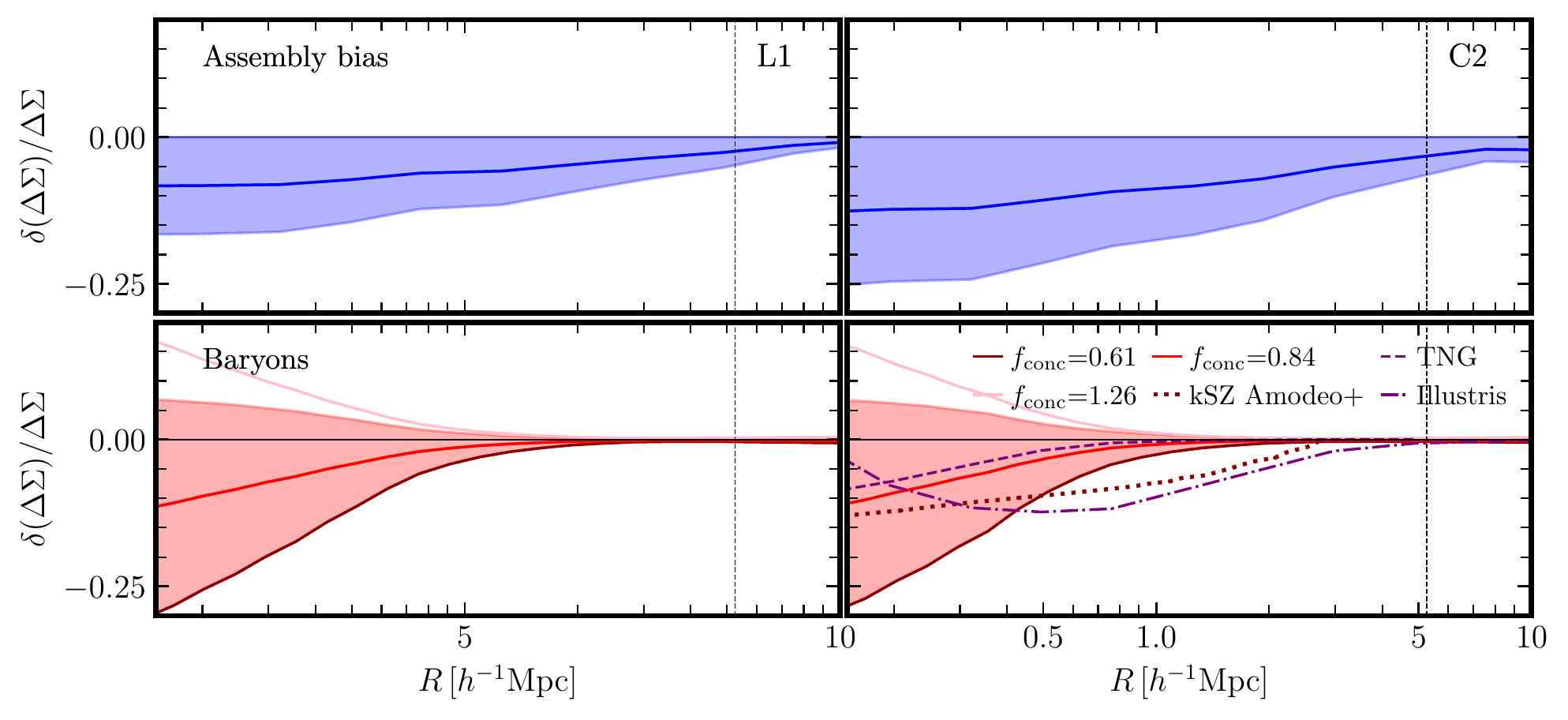}
\caption{The potential contribution from model uncertainties to the lensing signal as a function of $R$ for both the lowest redshift lens bin (L1, left-hand side) and the highest redshift bin (C2, right-hand side). Each panel shows the fractional contribution, derived from a model variant that accounts for one systematic, $\Delta\Sigma_{\rm sys}$, relative to the fiducial model, $\Delta\Sigma_{\rm fid}$, as $\delta(\Delta\Sigma)/\Delta\Sigma=(\Delta\Sigma_{\rm sys}-\Delta\Sigma_{\rm fid})/\Delta\Sigma_{\rm fid}$. The fractional impact is estimated for: 1. assembly bias (top panel), estimated following \citet{Yuan:2021}, using the \textsc{AbacusHOD} framework and including a secondary dependency on the local environment; 2. baryonic feedback (bottom panel), which we account for by adjusting the concentration normalisation parameter, $f_{\rm conc}$, in an NFW framework, based upon findings in \citet{Debackere2020}, with values drawn from the posterior of \citet{Viola:2015}. 
These are compared to previously estimated predictions based on hydrodynamic simulations, TNG and Illustrius \citep[purple dashed and dot-dashed;][]{Lange2019}, and based on kSZ measurements from \citet{amodeo21} (dotted). Similar behaviour is seen for lens redshfit bins, L2 and C1. In each panel the solid line is our fiducial estimate for the bias due to each effect and the shaded region is the corresponding error budget for that estimate. Note that while assembly bias and baryons bias the lensing signal low, as accurate modeling of these effects is not yet possible, our error budget encompasses the zero-impact case within $1\sigma$.}
\label{fig:eacherror}
\end{figure*}

\section{Small-scale systematics}\label{sec:uncertainties}

On small scales, the applicability of the HOD method in the presence of assembly bias and complex galaxy selections is still under investigation \citep[e.g.][]{Zentner:2014}. Additionally, the impact of baryonic feedback on the mass and galaxy distributions in group-sized haloes, at an intermediate redshift, is poorly understood \citep[e.g.][]{vandaalen2014}. In this section, we estimate the impact of these known sources of systematics on the GGL signal.  

\subsection{Baryonic Effects}

Baryon feedback alters the matter distribution \citep[see][for a comprehensive review]{chisari2019} and is therefore expected to impact both the galaxy--galaxy lensing and clustering signals \citep[e.g.][]{vandaalen2014, renneby20, van-Daalen:2020}. However, because AGN feedback can change the galaxy and halo distribution in different ways, the impact of baryons on clustering and GGL measurements is not necessarily the same.

As shown with hydrodynamical simulations, galaxy and halo properties can be altered by AGN feedback, impacting the stellar mass--halo mass relation. For example, in the model of \citet{vandaalen2014}, a higher halo mass corresponds to a lower stellar mass than in a DM-only simulation. In a similar way, this can modify the stellar masses of satellites and therefore the small-scale clustering. In the approach of using the clustering fits to predict the lensing, this baryonic effect can be captured by the halo model description, if the HOD parameterisation provides sufficient freedom to capture this. In addition, the spatial distribution of satellites can be changed through the back-reaction of feedback processes on the distribution of dark matter \citep[e.g.][]{van-Daalen:2011}, which alters both the clustering and GGL measurements. This is effectively a change in the concentration for the satellite distribution relative to the dark matter distribution. 

On the other hand, as GGL measures the projected matter density, it is sensitive to the change in the overall matter distribution due to the redistribution from baryons \citep[e.g.][]{schneider2019quantifying, Debackere2020} but this information does not enter the halo model for the clustering measurements. To first order, this effect can be accounted for in a halo model prescription by simply adjusting the concentration normalisation, denoted as $f_{\rm conc}$, in the Navarro-Frenk-White density profile \citep[NFW; ][]{Navarro:1997}. A change in the normalisation of concentration--mass relation has been commonly used in the literature \citep{van_den_Bosch_2013, Cacciato:2013, Viola:2015, Dvornik2018}, and has been shown to account for the baryonic feedback although these relations are calibrated on dark matter-only simulations \citep{Zentner2008, Debackere2020}. This is motivated by the fact that the AGN feedback pushes the baryons and dark matter from halo centres to their outskirts, effectively changing the concentration of the matter distribution \citep{McCarthy2017, vogelsberger14, pillepich18, Debackere2020, HMcode}. The direction of this effect is supported by the data \citep{Viola:2015}, which shows that the values of concentration normalisation prefer a value closer to 0.8 than 1.0, and by the fact that the halo masses are in agreement with hydrodynamic simulations that contain AGN feedback. 

In order to capture the impact of baryonic effects, we make predictions for the lensing signal with several values of the concentration normalisation parameter. 
That is, we alter the normalisation of the concentration--mass relation of the matter NFW profile (in $P_{\rm hm}$, entering the galaxy--galaxy lensing power spectra through equation~\ref{eqn:pgm}). We adopt the HOD posteriors from a large-scale joint fit of the clustering and lensing (see Section~\ref{sec:joint}) using a \textit{Planck} cosmology and explore the extent of the impact for $f_{\rm conc}$=0.84 ($\Delta\Sigma_{\rm fconc=0.84}$), 0.61, and 1.26 compared to 1.0, based on the best and 1$\sigma$ constraint from \citet{Viola:2015} for GAMA groups, which have the same stellar mass range as our BOSS galaxies \citep{Parejko:2013}. Figure~\ref{fig:eacherror} shows the fractional contribution, $\delta(\Delta\Sigma)/\Delta\Sigma$ of the baryonic feedback for bins L1 and C2, derived from model variants that account for it, $\Delta\Sigma_{\rm sys}$, relative to the fiducial model.
We find that while the clustering signal remains unchanged, this effectively reduces the GGL signal amplitude in the one-halo regime. An additional effect is that the feedback changes the halo mass with respect to the dark matter only case, which is the mass function provided by the {\sc Dark Emulator}. This would impact the two-halo term since the masses of all halos are altered, due to the change in density profiles. To incorporate this additional subtle effect consistently would require a more complex procedure. 

We compare this halo model estimate of the impact of baryon feedback to hydrodynamical simulations in the literature. Using the Illustris \citep{vogelsberger14} (purple, dot-dashed) and TNG-300 simulations \citep{pillepich18} (purple, dashed), \citet{Lange2019} measured the lensing signal around intermediate stellar mass (log$ M_*=10.5-12$) haloes, comparing the dark matter only case to that with baryons. As shown in the lower panel of Figure~\ref{fig:eacherror}, Illustris exhibits a greater suppression due to baryonic feedback that peaks at intermediate scales: TNG-300 sees a maximal effect of 10\% at $1h^{-1}$Mpc, similar to the $f_{\rm conc}$ estimate, while Illustris  sees an effect of 15\% at $3h^{-1}$Mpc. This difference arises from the fact that Illustris and TNG-300 have different (subgrid) implementations of the AGN feedback mechanism, with the TNG-300 matching observed galaxy and IntraCluster Medium properties more closely \citep{weinberger17,springel2018}. While \citet{Lange2019} only considered haloes representative of the CMASS sample ($z_{\rm l}=0.55$), \citet{van-Daalen:2020} explored the range of predicted baryonic effects exhibited across existing hydrodynamical simulations and found that the small-scale suppression of power due to baryonic feedback increases at low redshift. Furthermore, Illustris over-predicts the effect of feedback on the matter power spectrum due to its too-low baryon fraction in $\sim10^{14}\Msol$ haloes, while IllustrisTNG under-predicts this impact due to their too-high baryon fraction at the same mass scale \citep{van-Daalen:2020}. The baryon correction estimated in \citet{amodeo21} (dark red, dashed) using measurements of the kinematic Sunyaev-Zeldovich (kSZ) effect, was shown to account for 50\% of the discrepancy between clustering and lensing shown in \citet{Leauthaud:2017aa}. This impact corresponds to a reduction in the signal of 20\% at scales below 1$ h^{-1}$Mpc and larger than that found in \citet{Lange2019} using the TNG300 simulations or in this work.

It is evident that the extent of baryonic effects is uncertain. For our fiducial estimate of the correction due to baryonic feedback effects, $A_{\rm bary}$, we assume the $f_{\rm conc}$=0.84 variant, $\Delta\Sigma_{\rm fconc=0.84}$, the impact of which is indicated by the red line. For the uncertainty on this correction, $\sigma_{\rm bary}$, we assume that it is symmetric with an amplitude reflecting the difference between $f_{\rm conc}$=0.84 and 0.61, indicated by the red shaded region in Figure~\ref{fig:eacherror}. This error budget is sufficiently broad to encompass a null hypothesis corresponding to no baryon feedback.

\subsection{Assembly bias}

An assumption inherent to the modelling framework in Section~\ref{sec:th} is that dark matter halo mass is the only variable governing the occupation of haloes with galaxies.  If dark matter halo mass were the only variable determining the clustering of haloes, this simplification would not impact the predictions for $\Delta\Sigma$ at fixed $w_{\rm p}$. However, it has been shown that any dependency on secondary halo properties other than mass can significantly impact the galaxy clustering and lensing prediction \citep[e.g.][]{Gao:2005, Wechsler:2006, Gao:2007, Leauthaud:2017aa,wechsler2018, Hadzhiyska:2020, Yuan:2021}, an effect called assembly bias. This systematic primarily impacts small-scale measurements, but can also have a non-negligible impact on intermediate scales \citep[][]{Sunayama:2016, Yuan:2021}. 
 
It is important to note that the assembly bias, or secondary bias more generally, is the combination of two effects. One effect is the variation in galaxy--halo connection due to secondary halo properties, specifically termed galaxy assembly bias \citep{wechsler2018}. The second effect is the dependency of halo clustering on secondary halo properties other than mass, known as halo assembly bias \citep[e.g.][]{Croton:2007, Mao:2018}. The interplay of the two effects is what makes assembly bias important. 

Specifically, if there is no halo assembly bias, i.e. if halo clustering only depends on halo mass, then galaxy assembly bias would not actually contribute to two-halo clustering. In a simulation-based model, the halo assembly bias should be automatically accounted for in the N-body evolution, so the only piece that needs to be modeled is the galaxy assembly bias, hence the need for extended HOD models.

To calculate the model uncertainty due to assembly bias we use the \textsc{AbacusHOD} framework. In Appendix~\ref{HODimpl} we assess the model variance between this framework and our fiducial {\textsc Dark Emulator} when fitting the  projected two-point galaxy clustering for HOD parameters and predicting a lensing signal. Following the approach elaborated in \citet{Yuan:2021b}, we compute two GGL predictions following \citet{Yuan:2021}: The first represents the vanilla five-parameter HOD (plus incompleteness), tuned to match the observed projected two-point galaxy clustering in CMASS and LOWZ (as described in Appendix~\ref{HODimpl} and for the second, we extend the vanilla HOD model to include velocity bias and a secondary dependency on the local environment, tuned the extended model to the small-scale redshift-space clustering\footnote{Specifically, we choose $\xi(R, Z)$ as the small-scale redshift-space clustering data vector, with eight logarithmically spaced bins between 0.169$h^{-1}$Mpc and 30$h^{-1}$Mpc in the transverse direction, and six linearly spaced bins between 0 and 30$h^{-1}$Mpc bins along the line-of-sight direction.}. Note that the change to the HOD-based lensing prediction due to fitting redshift-space clustering (instead of $w_{\rm p}$) is small relative to the effect of assembly bias, as shown in \citet{Yuan:2021b}. 
Velocity bias \citep{Guo:2015a} is necessary to model the small-scale velocity signatures in the redshift-space clustering. The environment-based assembly bias is included as a result of both clustering analysis \citep{Yuan:2021b} and hydrodynamical simulation based studies \citep{Hadzhiyska:2020}, which found the environment to be the necessary secondary halo property in order to account for galaxy assembly biases, defined as the overdensity of dark matter subhaloes beyond the halo radius but within a $5h^{-1}$Mpc radius\footnote{The choice of $5h^{-1}$Mpc is motivated by internal tests done on hydrodynamical simulations and those carried out in \citet{Yuan:2021}, where we found $5h^{-1}$Mpc to provide the best fit on CMASS data.}. The final \textsc{AbacusHOD} decorated HOD model consists of ten parameters: five parameters from the vanilla HOD, the two velocity bias parameters, the two environment-based secondary bias parameters, and the incompleteness factor. These ten parameters are optimized to match the observed redshift-space clustering, $\xi(r_p, \pi)$, on small scales and the observed number density. See \citet{Yuan:2021b} for details of these fits and predictions.

We show in the second panel of Figure~\ref{fig:eacherror} that taking into account the impact of assembly bias suppresses the GGL signal, by an amplitude $A_{\rm ab}$, and we assume the difference between the two GGL predictions (blue shaded region) as a symmetric estimate of the model uncertainty due to secondary biases, such that $\sigma_{\rm ab}=A_{\rm ab}$. Thus, this error budget encompasses no assembly bias.

\subsection{Unmodelled systematics}
Here we discuss systematics that are not considered in the analysis and their plausible contribution to the results.

In the absence of lensing, galaxies are not randomly oriented. On large scales, galaxy shapes are influenced by the tidal field of the large-scale structure, while on small scales, effects such as the radial orbit of a galaxy in a cluster can affect their orientation. The correlation of the shape with the density field is referred to as `intrinsic alignment' \citep[IA; e.g.][]{Blazek_2012,troxel15, joachimi2015}. For high stellar mass elliptical galaxies like our BOSS sample, this systematic needs to be considered. For GGL, it is the impact of radial alignments that dominate the IA signal, which arise due to over-densities in the lens sample and a physically associated source sample.
Overall, the boost factor, introduced in Section~\ref{sec:measurements}, is a reasonable gauge for the extent of the clustered lens--source overlap, at a given radius. 
In this paper, we attempt to minimise the boost factor, and therefore mitigate the effect of intrinsic alignments by selecting source galaxies whose redshifts are sufficiently separated along the line of sight from the lenses. 

The boost factors are less than 15\% at the smallest scales and negligible at larger scales, from which we can approximate that, to first order at most 15\% of sources at small scales are associated with the lens sample. Furthermore, \citet{fortuna20} showed that the source samples' red-fraction (which have the highest contribution to intrinsic alignments) is low and that the IA signal decreases with the distance from the lens, such that the fraction of galaxies that could be affected by alignment would be even smaller than the source galaxy sample used. Applying these contamination fractions to the estimates for one--halo \citep{georgiou2019, fortuna20} and two--halo \citep{Singh:2017aa} intrinsic alignment contributions, we argue that this systematic has a negligible impact on this analysis.

The central and satellite components of the halo model are described here simply in terms of one- and two-halo terms describing the halo component and the large-scale structure component. They do not include additional terms of the halo model, such as the effect of satellite stripping or other impacts in the transition regime. 
These are less dominant compared to the contribution from the satellites for the lensing signals $\sim R > 0.1 {\rm Mpc}/h$ \citep[see, for example, figure 5 in][]{y3-HOD}, and so we regard it as unimportant in our analysis, except for the smallest radial bin.

\section{Measurements}\label{sec:measurements}

In this section, we describe the estimators for the excess surface mass density in Section~\ref{sec:ds}, using the cross correlation of the spectroscopic lens sample and multiple sources of background lensing data and the projected galaxy clustering in Section~\ref{sec:wp}. 

\subsection{Galaxy--galaxy lensing}\label{sec:ds}

The excess surface density can be related to the average tangential shear $\langle \gamma_{\rm t} (\theta) \rangle$ as
\begin{equation}
\Delta \Sigma = \frac{\langle \gamma_{\rm t} (\theta) \rangle}{\overline{\Sigma_{\rm c}^{-1}}} \, 
\end{equation}
at a projected separation $\theta=R/\chi(z_{\rm l})$, where $\chi(z_{\rm l})$ is the comoving distance to the lens. \amon{We adopt a flat $\Lambda$CDM cosmology with $\Omega_{\rm m}=0.3$ and $H_0=0.7$ km s$^{-1}$ Mpc$^{-1}$ when computing comoving distances.}

For a source redshift distribution $n(z_s)$ the average inverse critical density is given by
\begin{equation}
\overline{\Sigma_{\rm c}^{-1}}(z_{\rm l}) = \frac{4\pi G (1+z_{\rm l}) \chi(z_{\rm l})}{c^2}\int_{z_{\rm l}}^\infty \mathrm{d}z_{\rm s} n(z_{\rm s}) \frac{\chi(z_{\rm l},z_{\rm s})}{\chi(z_{\rm s})} \, ,
\end{equation}
where the source redshift distribution is computed for a given lens redshift $z_l$ and normalised such that $\int_0^{\infty} n(z_s) dz_s = 1$. 

An estimator for the `stacked' excess surface density for a sample of lens galaxies can therefore be written as
\begin{equation}
\Delta \Sigma_{\rm l} (R) = \frac{1}{1+\overline{m}} \frac{\sum_{\rm ls} w_{\rm ls} \epsilon_{\rm t} (\overline{\Sigma_{\rm c}^{-1}})^{-1}(z_{\rm l})}{\sum_{\rm ls} w_{\rm ls}} \, ,
\end{equation}
where the summation is over all lens--source pairs in a given radial bin defined by $R$ and $\epsilon_{\rm t}$ indicates the tangential ellipticity of the source. Note that for KiDS, $\overline{\Sigma}_{\rm c}(z_{\rm l})$ is computed per lens, with the entire ensemble of sources, whereas for DES, it is computed per lens bin, where $z_{\rm l}$ is the mean redshift. This approximation is negligible. The parameter $\overline{m}$ is the multiplicative bias correction and $w_{\rm ls}$ is the weight assigned to lens--source pair ${\rm ls}$ given by
\begin{equation}
w_{\rm ls} = W_{\rm s} w_{\rm l} \, ,
\label{eqn:sw} 
\end{equation}
where $ W_{\rm s}$ is the source weight defined for each survey. 

We subtract the excess surface density measured around random positions within the survey footprint, $\Delta \Sigma_{\rm r}$, which is predicted to be zero, on average, in the absence of an additive bias \citep{Mandelbaum_2005, Mandelbaum_2013, Singh_2017}. Finally, due to uncertainties associated with photometric redshifts, the selection of source galaxies to be behind a lens is imperfect, resulting in a biased estimate of the excess surface density. This can be corrected for by applying a boost factor, $B(R)$, computed using un-clustered, random positions that have the same selection as the lens galaxies. The boost factor is given by 
\begin{equation}
B(R) = \frac{\sum_{\rm ls} w_{\rm l} w_{\rm s}}{\sum_{\rm rs} w_{\rm r} w_{\rm s}} \, ,
\label{eqn:boost}
\end{equation}
where  $w_{\rm l}$ and $w_{\rm r}$ corresponds to any weighting applied to the lens galaxies or random positions respectively, and are normalised such that  $\sum_{\rm l} w_{\rm l} = \sum_{\rm r} w_{\rm r}$. 
Finally, the overall estimator is
\begin{equation}
  \Delta \Sigma(R) = B(R) \Delta \Sigma_{\rm l} (R) - \Delta \Sigma_{\rm r} (R) \, .
\label{eqn:dseqn}
\end{equation}

\subsubsection{DES Y3}
The DES Y3 $\Delta\Sigma(R)$ measurements are computed following the methodology outlined in the previous section. The analysis setup used in these calculations is publicly available in the \textsc{xpipe} package\footnote{\url{https://github.com/vargatn/xpipe}}. To estimate the error, a Jackknife covariance is computed by splitting the lens
galaxies and random points into 75 regions using the \textsc{kmeans} algorithm. For a footprint area of $\sim770$ deg$^2$, this yields regions of approximately 3 deg$^2$ on a side assuming a square geometry, or $\sim65h^{-1}$Mpc (and the largest angular scale
we measure is $\sim$50$h^{-1}$Mpc). These measurements are shown for each lens bin in Figure~\ref{fig:ds}.

The source weighting, $W_{\rm s}$ is an inverse-variance weight, described in \citet{y3-shapecatalog}. In all other respects, the estimator for each DES source bin and BOSS lens bin is equivalent to equation~\ref{eqn:dseqn} with an additional factor for the \textsc{metacalibration}-derived weights, $1/\left<R\right>=1/\left(\mathsf{R}^{\rm T}_{\gamma,s}  + \langle \mathsf{R}^{\rm T}_\mathrm{sel} \rangle \right)$. The \textsc{metacalibration} algorithm \citep{Huff2017,Sheldon2017} provides estimates on the ellipticity, $\epsilon_{\rm t},$ of galaxies, the \emph{response} of the ellipticity estimate on shear, $\mathsf{R}_{\gamma,s}$, and of the ensemble mean ellipticity on shear-dependent selection, $\mathsf{R}_\mathrm{sel}$. These are applied in the shear estimator to correct for the bias of the mean ellipticity estimates. 
The DES shear response is broken into two terms: $\mathsf{R}^{\rm T}_{\gamma,s}$ is the average of the shear responses measured for individual galaxies including selection effects and $\langle \mathsf{R}^{\rm T}_\mathrm{sel} \rangle$ is the response of the selection effects to a shear. Each galaxy has a unique value for $\mathsf{R}^{\rm T}_{\gamma,s}$, while $\langle \mathsf{R}^{\rm T}_\mathrm{sel} \rangle$ is a single number computed for each source galaxy ensemble. 

The DES Y3 multiplicative bias correction factors, $\overline{m}$, are provided for each redshift bin \citep{y3-imagesims} and are applied directly to the data vector. This differs from the various DES analyses in that in those analyses the systematic uncertainties were incorporated at the model/likelihood level, and their amplitudes varied according to their respective prior. 

A boost factor is estimated following equation~\ref{eqn:boost}. It is validated against an alternative method, estimated using $p(z)$ decomposition \citep{Gruen2014,Varga2019} that is designed to minimize potential spatial variations in the shear selection performance (a potential issue with bright galaxies).  This correction is small: at scales greater than $1h^{-1}\mathrm{Mpc}$ the boost factor is negligible, and therefore does not impact our large scale measurements. At the smallest scale measured, it is at most a 15\% effect. 

The systematic uncertainties in the DES Y3 redshift and shear calibration are accounted for as follows.
For each of the four tomographic source bins, the mean lensing source redshift distribution is calibrated in \citet{y3-sompz} as the spread of 6000 realisations of the source redshift distributions. In \citet{y3-imagesims}, these are modified to include the uncertainty due to the shear calibration, to describe the joint systematic uncertainties. For each lens bin in our sample and the corresponding source bins used, we compute the mean lensing efficiency, $\langle\eta\rangle$, for each of the realisations. Following the measurement, the source bins are combined using $\Sigma_\mathrm{crit}^{-1}$ weights and the standard deviation across the realizations of $\langle\eta\rangle$ are extracted for each lens bin as the systematic uncertainty. These are rounded up as $\sigma_{\rm sys}$=[1\%, 1.5\%, 2\%, 2.5\%], and combined with the statistical Jackknife uncertainty.

\subsubsection{KiDS-1000}
The KiDS lensing signal is computed similarly to the methodology outlined in \citet[][]{Dvornik2018} and \citet[][]{Amon2018}. Errors are computed using a bootstrap method using regions of 4 ${\rm deg}^2$, or $\sim$80$h^{-1}$Mpc (the largest angular scale
we measure is $\sim$50$h^{-1}$Mpc). 

The source weighting from equation~\ref{eqn:sw} is defined as
\begin{equation}
W_{\rm s} = w_{\rm s} \Big[\overline{\Sigma}_c^{-1}(z_{\rm l})\Big]^2 \, ,
\end{equation}
where $w_{\rm s}$ is the per-galaxy \textit{lensfit} weight \citep{Miller2013}, which approximately corresponds to an inverse variance weighting, $w^{-1} \sim \sigma_e^2 + \sigma_{\mathrm{rms}}^2$. 

The multiplicative shear calibration correction \citep[][]{kannawadi19} is estimated for the ensemble source and lens galaxy population. The multiplicative bias has been estimated for a set of tomographic bins described in \citet{asgari20}. These corrections are then optimally weighted and stacked following 
\begin{equation}
\overline{m} = \frac{\sum_{\rm ls} w_{\rm ls} m_{\rm s}}{\sum_{\rm ls} w_{\rm ls}} \, ,
\label{eqn:mbias}
\end{equation}
where $m$ is the multiplicative bias value for the tomographic bin that source galaxy $s$ falls into given its $z_\mathrm{B}$ value. The resulting correction to the measurement for each lens sample is  $\overline{m}=\{-0.001,0.002,0.004,0.005\}$, which is independent of the distance from the lens, and reduces the effects of multiplicative bias \citep[][]{kannawadi19}.

The lensing signal around lens galaxies is computed following the equations presented in Section~\ref{sec:ds}. Source galaxy contamination is accounted for by including a boost factor, such that the uncertainty is also inflated by this factor. This correction is small: at scales greater than $1h^{-1}\mathrm{Mpc}$ the boost factor is negligible, and at the smallest scale measured, it is at most a 15\% effect. 

We estimate the systematic uncertainty due to the errors in the redshift distribution. The shift in the mean of the redshift distribution, $\Delta z$, was determined for five tomographic bins in \citet{SOM,Hildebrandt2021}. From these values we estimate the weighted average $\Delta z$ per lens bin, $\overline{\Delta z} = \sum_{\rm ls} w_{\rm ls} \Delta z_{\rm s}/\sum_{\rm ls} w_{\rm ls}$, where $\Delta z_{\rm s}$ is assigned to each source depending on which tomographic bin it falls into given its $z_\mathrm{B}$ value. We then remeasure our lensing signal with the source redshift distributions shifted by the $\overline{\Delta z}$ for lens sample. The resulting change in amplitude of the lensing signal is averaged over all scales and taken to be the systematic error due to the uncertainty in the photometric redshift distributions, which we find to be up to $1 \%$ for all lens bins.  We follow the same procedure for estimating the impact of the uncertainty on the multiplicative bias, which we also find to be less than $1\%$. Combining these two uncertainties in quadrature gives an overall systematic error of less than 1.5\%, which we combine with the statistical error. 

\subsubsection{HSC Y1}
The HSC measurements for $\Delta\Sigma(R)$ are described and presented in \citet*[Section 6.3][]{LWB}. We briefly note two main differences in the methodology used to construct these measurements compared to that of KiDS and DES. First, per-galaxy, point-estimate redshifts are used in the computation, opposed to a calibrated distribution of the ensemble. Second, two redshift cuts are used to define the source sample behind the lens along the line of sight: $z_{\rm s} > z_{\rm l} + 0.1$ and $z_{\rm s} > z_{\rm l} + \sigma_{68}$, where $\sigma_{68}$ is the 1$\sigma$ confidence limit of the photometric redshifts, and no boost factor is applied. Here we combine in quadrature the statistical error with the systematic uncertainty, the latter of which is reported in \citet*{LWB} as 5\%.

\subsubsection{Magnification Bias}\label{sec:lens_mag}

In addition to the effect of shear, the solid angle spanned by a galaxy image is modified by a magnification factor compared to the solid angle covered by the galaxy itself. This effect alters the number density of the sample for a given set of cuts. Lens magnification describes this effect on the number density of the lens galaxy sample by intervening structure. Following \citet{Elvin2022}, in this analysis we consider only the effect on the lens galaxies, which is dominant compared to source magnification for GGL \citep{Mandelbaum:2006b}. Two mechanisms are at work: at higher lens redshift, more intervening matter is present, such that the impact of magnification effects grows with increasing redshift. On the other hand, the impact is reduced with increasing line-of-sight separations of lenses and sources. 

To account for the impact on the lensing signals, the magnification angular power spectrum, $C^{\rm gm}_{\rm lmag}(\ell),$ is computed by integrating the intervening matter up to the lens redshift \citet{Elvin2022}. The contribution of lens magnification to the tangential shear is determined as
\begin{equation}
{\gamma_t}_{\rm lmag}(\theta) = 2 (\alpha_{\rm lmag} -1) \int  \frac{\ell d\ell}{2\pi}C^{\rm gm}_{\rm lmag}(\ell)J_2(\ell\theta) \, ,
\end{equation}
where $\alpha_{\rm lmag}$ depends on the properties of the lens sample. The magnification power spectrum is then defined as 
\begin{multline}
C^{\rm gm}_{\rm lmag}(\ell) = \int d\chi \frac{g_{\rm g}(\chi) \, g_{\rm s}(\chi)}{\chi^2} P^{\rm nl}_{\rm m}\Big(\frac{\ell+1/2}{\chi},z(\chi)\Big) \, ,
\label{eqn:mag}
\end{multline}
where the lensing window function of the galaxies $g_{X}$, for $X={s,g}$ is defined as
\begin{equation}
g_{X}(\chi) = \frac{3 \Omega_{m} \chi }{ 2c^2 [1 + z(\chi)] } \int_{0}^{\chi} d\chi^{\prime} n_{\rm X}(\chi^{\prime}) \frac{dz/d\chi^{\prime}}{c}
\frac{\chi - \chi^{\prime}}{ \chi^{\prime}} \,.
\end{equation}

For this analysis, following \citet{mag2021}, we define $\alpha_{\rm lmag}={1.93 \pm 0.05}$ for LOWZ and $\alpha_{\rm lmag}={2.62 \pm 0.28}$ for CMASS. Figure 7 in that work shows only a slowly varying redshift dependence. Given that this correction is small, we ignore the redshift dependence across each of the BOSS samples and assume the same value for L1 and L2, as well as for C1 and C2. We correct the lensing signals for this sub-dominant systematic and neglect any uncertainty on the value of $\alpha_{\rm lmag}$, following \citet{joachimi20}. In Appendix~\ref{app:mag}, we demonstrate that the impact of magnification is small, but increases with redshift such that the correction is most significant for the C2 bin, still remaining less than $\sim10$\%. We note that magnification bias also impacts the clustering measurement, but we neglect this as it has been shown to be small \citep{thiele2019}.

\subsubsection{Combined signals: KiDS-1000 + DES-Y3}\label{sec:deskids}
Given that the KiDS--BOSS and DES--BOSS on-sky footprints have no overlap, we take the two measurements to be independent and assess their consistency (see Figure~\ref{fig:ds}). To do so, we adopt a model-independent approach: we compute the difference between the two lensing surveys' signals and fit to a null signal, which is the expectation for perfect agreement. This approach assumes that the data can be described by Gaussian likelihoods and are independent, such that the covariance of the difference is equivalent to the sum of that of the individual measurements.

The difference in signals compared to null has a p-value of $(0.49,0.20,0.82,0.66)$ and a reduced $\chi^2$ of $(1.0,1.4,0.70,0.86)$ for each redshift bin. A p-value less than 0.01 equates to a 99\% confidence in rejection of consistency between the data, assuming Gaussian statistics. We can convert these p-value estimates into the more intuitive quantity of number of $\sigma$; for this we find $(0.69,1.3,0.23,0.45)$. Our p-value is always larger than 0.2, so we conclude that our two sets of GGL measurements from KiDS-1000 and DES-Y3 are statistically consistent. 
As such, we compute a combined DESY3+KiDS1000 measurement by taking the inverse-variance weighted average, shown as the green data points in Figure~\ref{fig:ds}. In Appendix~\ref{sec:lwb} we discuss the consistency of the DESY3+KiDS1000 measurement with HSC, shown in the same figure in yellow. We cannot combine KiDS or DES with HSC as the overlapping area between them is significant, as demonstrated in Figure~\ref{fig:map}.

\subsection{Projected clustering}\label{sec:wp}

We compute the projected correlation function, $w_{\rm p}$ using the three-dimensional positional information for each of the four spectroscopic lens samples over the entire BOSS area. We measure these statistics using random catalogues \citep{Reid:2016} that contain $N_{\rm ran}$ galaxies, roughly 40 times the size of the galaxy sample, $N_{\rm gal}$, with the same angular and redshift selection. To account for this difference, we assign each random point a weight of $N_{\rm gal}/N_{\rm ran}$. In addition, we neglect the redshift weight, $w_{\rm z}$, and use only the $w_{\rm systot}$ for the case of CMASS ($w=1$ for LOWZ). Instead, we account for spectroscopic incompleteness due to fibre collisions using the algorithm developed by \citet{Guo:2012aa}. 

Adopting a fiducial flat $\Lambda$CDM WMAP cosmology \citep{Komatsu:2009aa} with $\Omega_{\rm m}=0.3$, we estimate the 3D galaxy correlation function, $\xi_{\rm gg}(R,Z)$, as a function of comoving projected separation, $R$, and line-of-sight separation, $Z$, using the estimator proposed by \citet{Landy:1993},
\begin{equation}
\xi_{\rm gg}(R,Z)=\rm{\frac{dd-2dr+rr}{rr}}\, ,
\end{equation}
where $\rm{dd, rr}$ and $\rm{dr}$ denote the weighted number of pairs with a separation $(R,Z)$, where both objects are either in the galaxy catalogue, the random catalogue or one in each of the catalogues, respectively.

In order to obtain the projected correlation function, we combine the line-of-sight information by summing over 50 linearly spaced bins in $Z$ from $Z=0$ to $Z=100 \, h^{-1}$Mpc \citep{Guo2018}, 
\begin{equation}
w_{\rm p}(R)=2 \sum_i \xi_{\rm gg}(R, Z_i) \Delta Z_i \, .
\end{equation}

We use 17 logarithmic bins in $R$ from $R=0.05$ to $R=50 \, h^{-1}$Mpc. The upper bound $Z_{\rm max}=100 \, h^{-1}$Mpc can potentially create a systematic error as $R$ approaches $Z_{\rm max}$ due to any lost signal in the range $Z>100 \, h^{-1}$Mpc, however the signal is negligible on these scales and the measurement was robust to changes in this value for the level of precision of the analysis. The error in $w_{\rm p}(R)$ is determined via a Jackknife analysis, dividing the galaxy survey into 400 regions, ensuring a consistent shape and number of galaxies in each region.

\section{Large-scale lensing and clustering fits}\label{sec:joint}

\begin{figure*}
\centering
\includegraphics[width=\textwidth]{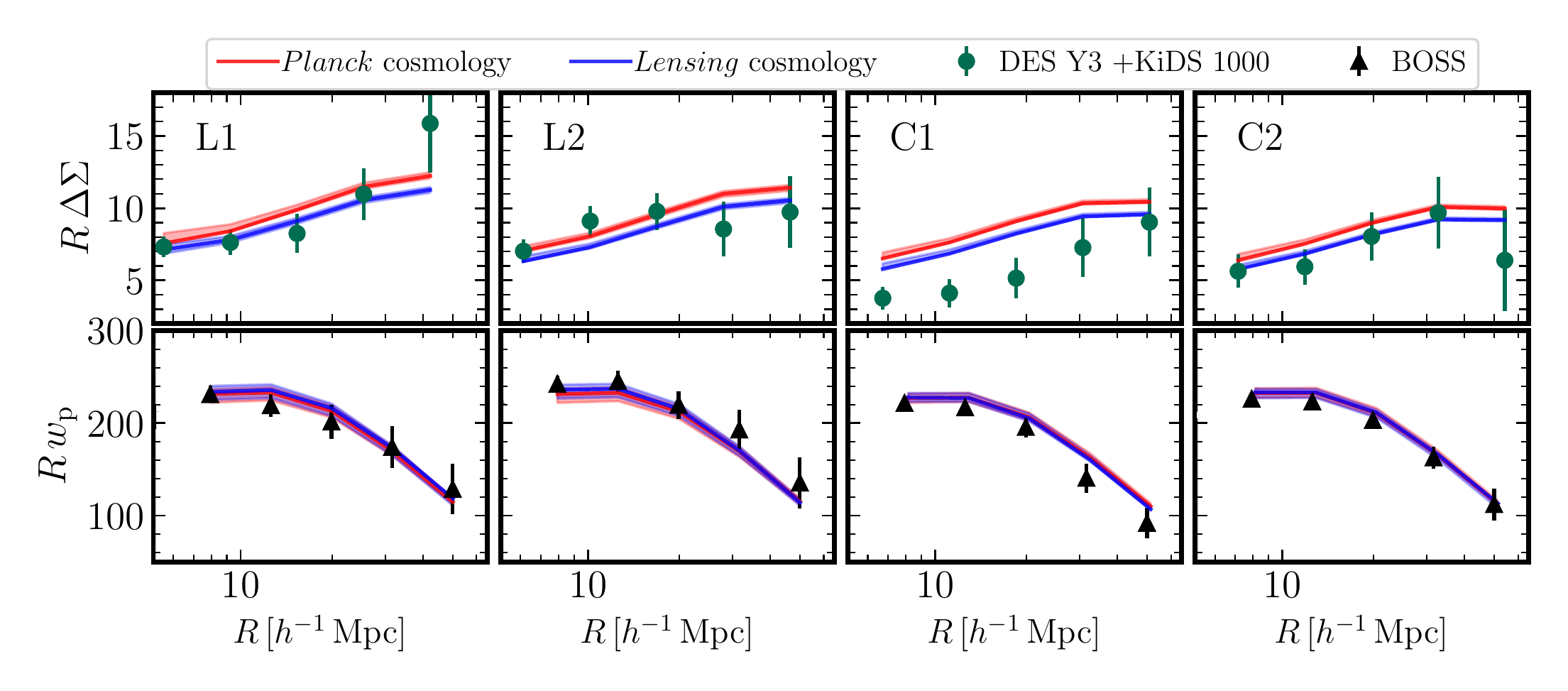}
\caption{Joint fits to the large-scale ($R>5.25h^{-1}{\rm Mpc}$) BOSS clustering measurements (black data points, lower panels), $w_{\rm p}$, and lensing profiles, $\Delta\Sigma(R)$, measured by DES Year 3 + KiDS-1000 (green, upper panels), computed using the {\sc Dark Emulator}, for each of the four lens samples. The fits are performed at a fixed cosmology, using parameters from \citet{Planck2018} (red), as well as a \textit{Lensing} cosmology (blue), using a lower value of $S_8$ \citep{Heymans2021}. The lines indicate the best-fit, and the corresponding shaded regions represent the 68\% confidence level derived from the posterior of each of the fits. For the clustering, the fit is largely independent to this change in cosmological parameters, while the difference is seen for the lensing. When limited to the large scales, the \textit{Lensing} cosmology is preferred, although the difference between the fits is not significant.}
\label{fig:joint}
\end{figure*}

To date, joint analyses from DES \citep{y3-3x2ptkp}, HSC \citep{HSC2x2}, and KiDS \citep{Heymans2021} have focused on either marginalising over the small-scale modeling systematics, such as those detailed in Section~\ref{sec:uncertainties}, or limiting the scales of the measurements used, as these limit the robustness of cosmological information from those scales. Following that, we first consider a joint fit of the large-scale clustering and DES+KiDS lensing measurements. We focus here on DES+KiDS only, as the HSC Year 1 measurements are limited to $R<15h^{-1}$Mpc, and so do not have sufficient signal-to-noise on large scales.
Guided by the potential impact of these systematics demonstrated in Figure~\ref{fig:eacherror}, we consider both the GGL and clustering measurements limited to R$>5.25h^{-1}$Mpc. 

We fit the galaxy--halo connection model described in Section~\ref{sec:th} to both the observed projected clustering of BOSS galaxies, $w_{\rm p}(R)$, and the excess surface mass density, $\Delta\Sigma(R)$ for each lens bin separately. As the clustering measurements were derived from an order-of-magnitude larger area than the lensing, any cross-covariance between the clustering and lensing is negligible \citep{More:2015, joachimi20}. The flat priors assumed for each parameter are reported in Table~\ref{tab:priors}, within a flat-geometry $\Lambda$CDM model, specified by the five cosmological parameters fixed at the values quoted in Table~\ref{tab:cosmo}, and with massive neutrinos with a fixed total mass of 0.06eV. A multi-variate Gaussian likelihood is used\footnote{We use the nested sampling \citep{Skilling2004} code {\sc MultiNest} \citep{feroz09} to evaluate the posterior of galaxy--halo connection parameters. We use 1,000 live points, a sampling efficiency parameter of 0.8, and an evidence tolerance factor of 0.5.}.

When fitting a clustering signal, the difference between cosmological parameters assumed for measurements and models should be taken into account, because it affects the radial and angular distances. Note that for the flat-$\Lambda$CDM cosmology, only $\Omega_{\rm m}$ is relevant for this effect. We correct for this effect following the prescription of \cite{More:2013aa}. The detailed implementation is described in \cite{HSC2x2}. For our setup, where the difference is at most $\Delta\Omega_{\rm m}=0.015$ compared to the {\it Planck} cosmology, the correction factor is only a few percent. Such a correction is also considered for the lensing signal, but it is estimated to be sub-percent for our case, and as such, we ignore it in this study.

When assessing the goodness of fit, we evaluate the effective degrees of freedom using noisy mock data vectors. We first generate 30 noisy mock signals by deviating signals around the best-fit model following the covariance. We then perform the fit to each mock signal, make a histogram with the best-fit $\chi^2$ from mock analyses, and find the effective degree of freedom (DoF) by fitting a $\chi^2$ distribution. We need to rely on mock signals to derive the effective DoF because the posterior distributions of HOD parameters are highly correlated and have a strong non-Gaussianity and thus the Gaussian linear model by \cite{Raveri2019} is not valid for our large-scale measurements. For detailed discussions, see \citet{HSC2x2}.

\subsection{Considering a low-$S_8$ Universe}\label{sec:lowS8}

\begin{table}
		\centering
		\begin{tabular}{ccc}
			\hline
			 Parameter & \textit{Planck} Cosmology & \textit{Lensing} Cosmology\\
			\hline
			 $\omega_{\rm c} = \Omega_{\rm c}h^2$ & \multicolumn{2}{c}{0.120}  \\
			 $\omega_{\rm b} = \Omega_{\rm b}h^2$ & \multicolumn{2}{c}{0.022} \\
			 $n_{\rm s}$ & \multicolumn{2}{c}{0.965}\\
			 $\Omega_{\rm de}$ & 0.685 & 0.695\\
			 ${\rm ln}(10^{10}A_{\rm s})$ & 3.044 & 2.910\\
			 \hline
			 $S_8=\sigma_8(\Omega_{\rm m}/0.3)^{0.5}$ & 0.83 &  0.76 \\
			\hline
		\end{tabular}
		\caption{\label{tab:cosmo}The fixed values of the flat $\Lambda$CDM cosmological parameters defined for the \textit{Planck} cosmology \citep{Planck2018} (TTTEEE+lowE+lensing) and the \textit{Lensing} cosmology, corresponding to lower values of the $S_8$ parameter. The implementation of the {\sc Dark Emulator} used in this work assumes a flat-geometry $\Lambda$CDM model with massive neutrinos with a fixed total mass of 0.06eV, specified by the five parameters shown. $\Omega_{\rm c}$, $\Omega_{\rm b}$ and $\Omega_{\rm de}$ are the density parameters for cold dark matter, baryonic matter and dark energy, respectively. $A_{\rm s}$ and $n_{\rm s}$ are the amplitude and tilt parameters of the primordial curvature power spectrum normalized at $k_{\rm pivot} = 0.05$Mpc$^{-1}$. For the latter, cosmological parameters are taken from the combined KiDS-1000+BOSS+2dFLenS analysis \citep{Heymans2021}. Note that $\Omega_{\rm c}$,  $\Omega_{\rm b}$ and $n_{\rm s}$ are not varied between the two cases.}
	\end{table}

Given the mounting evidence that large-scale weak lensing analyses prefer lower values for the $S_8$ cosmological parameter than that constrained by the \citet{Planck2018} measurements of the primary CMB \citep[e.g.][]{y3-cosmicshear1,y3-cosmicshear2, asgari20, hikage19, Ham20}, it is interesting to compare these results within both cosmological frameworks. As such, we perform the fits at a fixed cosmology and compare the goodness of fit in two cases: the \citet{Planck2018} cosmology and a \textit{`Lensing cosmology'}, defined here. For this, we adopt the best-fit cosmology from the joint lensing and clustering analysis of \citet{Heymans2021}, which is consistent with the parameters inferred from DES \citep{y3-3x2ptkp}, as well as with cosmic shear constraints from DES, HSC and KiDS. Specifically, we update the dark energy density parameter, $\Omega_{\rm de}$, and the amplitude of the primordial curvature power spectrum, ${\rm {ln}}(10^{10}A_{\rm s})$, to those reported in Table~\ref{tab:cosmo}. Note that in {\sc Dark Emulator}, the matter density parameter, $\Omega_{\rm m}$ is defined via $\Omega_{\rm de}=1-\Omega_{\rm m}$, and the combination of $\omega_{\rm m}$ and $\omega_{\rm c}$ defines the dimensionless
Hubble parameter, $h$ via $h=[(\omega_{\rm m}+\omega_{\rm c})/(1-\Omega_{\rm m})]^{1/2}$.

The result of this comparison is shown in Figure~\ref{fig:joint}. The
p-values for the lens bins, by increasing redshift, are found to be [0.15,0.02,0.01,0.94] and [0.15,0.05,0.08,0.97]
for the \textit{Planck} and \textit{Lensing} cosmologies, respectively\footnote{There are five formal DoF (ten data points and five free parameters) that are reduced to [7.85,8.23,6.23,6.53] and [7.61,8.20,6.24,6.6.46], for the \textit{Planck} and \textit{Lensing} cosmologies, respectively, when the effective number is computed.}. When computing the $\chi^2$, we apply a \citet{Hartlap_2007} correction,\footnote{This is applied by replacing the likelihood, $L$, according to -log$(L)\Rightarrow-\alpha$log$(L)$, where $\alpha=(N_{\rm jk}-N_{\rm bin}-2)/(N_{\rm jk}-1)$, $N_{\rm jk}$ is the number of Jackknife patches and $N_{\rm bin}$ is the number of data bins used in the fit.} as the covariance matrix is estimated using a Jackknife method with a finite number of patches. It therefore is associated with it some measurement noise, such that $\hat{C}^{-1}$ is not an unbiased estimate of the true inverse covariance matrix. We compute the p-value from the measured $\chi^{2}$ value, assuming the effective DoF using mocks. We define acceptable goodness-of-fit as p-value$ \geq 0.01$ and find this to be acceptable for both cosmologies, with the exception of the C1 lens bin which is on the boundary of satisfying this criteria.
That is, limited by the current level of statistical power in the measurements on these large scales, we cannot distinguish between the two cosmologies with significance.

\vspace{0.5cm}
\section{On the consistency of clustering \& lensing}\label{sec:results}

Modern lensing surveys allow us to probe a rich set of physical processes, containing information on cosmology, galaxy formation, and feedback. Developing and testing a model powerful enough to explain the complexity of the data across all scales remains a work in progress for the community. For this reason many cosmological analyses have restricted their attention to the better-understood and more theoretically controlled large-scale clustering \citep[e.g.][]{Heymans2021, y3-3x2ptkp}. This mitigates biases in the inferred cosmological constraints that arise due to non-linear modelling systematics and the uncertain impact of baryonic feedback \citep{joachimi20, y3-generalmethods}. As reflected in Section~\ref{sec:joint}, there is insufficient statistical power with these data to give compelling evidence for either cosmology using the easier-to-model linear scales. On the other hand, small-scale measurements afford substantially more constraining power. An understanding of these scales is hindered by numerous hard-to-model physical effects of approximately comparable amplitude, demonstrated in Section~\ref{sec:uncertainties}. As these effects impact galaxy--galaxy lensing and galaxy clustering differently, an interesting avenue to understand them is to assess their consistency.

\subsection{Scaling amplitude}\label{sec:Alens}

We evaluate the consistency between the measurements by including an additional parameter, $A$, which multiplies the amplitude of the galaxy--galaxy lensing signal as
\begin{equation}
    \Delta \Sigma \rightarrow A\Delta\Sigma(R) \, ,
\end{equation}
allowing it to decouple from the model for the projected clustering. If the clustering and galaxy--galaxy lensing measurements are both well fit by the same model, we expect $A$ to be consistent with unity; any significant deviation implies that the measurements are not fully consistent within our chosen model. Here we use consistency of $2.5\sigma$ as the criterion, following \cite{y3-3x2ptkp,Heymans2021}.
Note that our approach of a joint fit using an inconsistency parameter differs from that of \citet{Leauthaud:2017aa, Lange2019, Lange2021}, where the fit to the BOSS clustering is used to predict the lensing signal. In those works, the level of inconsistency is then quantified as an averaged ratio between the predicted and observed GGL. On small scales, these probes are sensitive to complexities of the small-scale dark matter--galaxy connection, such as those described in Section~\ref{sec:uncertainties}.  To assess the scale dependence of the consistency, we consider this fit when isolating large and small scales.

\begin{figure*}
\centering
\includegraphics[width=\textwidth]{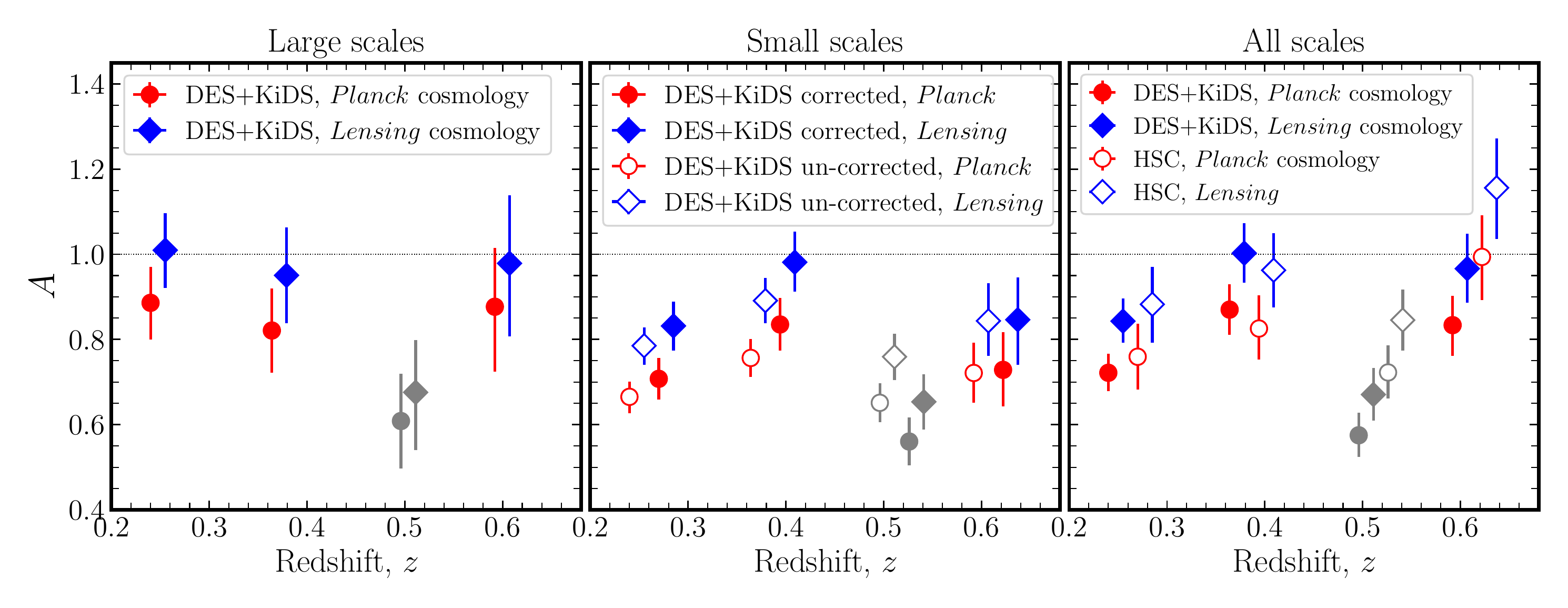}
\caption{Consistency of lensing and clustering measurements for each lens redshift bin, quantified by a consistency amplitude parameter, $A$,  that allows for a decoupling of the two measurements when $A$ deviates from $A=1$.  We report the marginal posterior of this parameter,  constrained in a joint fit of the two observables. Left panel: Large scales ($R>5.25h^{-1}{\rm Mpc}$)  with KiDS+DES, comparing the \textit{Planck} (red circles) and \textit{Lensing} cosmologies (blue diamonds). Middle panel: Small scales with KiDS+DES, showing the impact of accounting for modelling baryonic effects and assembly bias (filled markers), compared to neglecting them (empty markers), again for the two cosmologies. Right panel: All scales, showing comparison between KiDS+DES (filled markers) and HSC (empty markers), including the corrections for baryonic effects and assembly bias, again for the two cosmologies. The large-scale check fails for bin C1, with a constraint on $A$ that deviates from 1; we infer this as evidence for modelling systematics and denote the results for this bin with grey markers.}
\label{fig:alens}
\end{figure*}

\begin{table*}
\centering
\begin{tabular}{c|c|c|c|c|c|c|c|c}
	\hline
Cosmology & Scales & Data & \multicolumn{4}{c}{Inconsistency systematic parameter, $A$} \\
	 \hline
	
		 &  & & L1:$z$=0.15-0.31 & L2:$z$=0.31-0.43 & *C1:$z$=0.43-0.54 & C2:$z$=0.54-0.7 & All bins & All bins, no C1 \\
	\hline
    \textit{Planck}   & LS & DES+KiDS &  $0.89_{-0.09}^{+0.08}$  &   $0.82_{-0.10}^{+0.10}$ &   $0.61_{-0.11}^{+0.11}$ &    $0.88_{-0.15}^{+0.14}$  & $0.80_{-0.05}^{+0.05}$  & $0.86_{-0.06}^{+0.06}$ \, (2.3$\sigma$)\\
    & SS & DES+KiDS & $0.67_{-0.04}^{+0.04}$  &   $0.76_{-0.04}^{+0.04}$ &   $0.65_{-0.05}^{+0.05}$ &   $0.72_{-0.07}^{+0.07}$  &  $0.69_{-0.02}^{+0.02}$  & $0.71_{-0.03}^{+0.03}$  \\
   \textit{Planck-corr}   & SS & DES+KiDS & $0.67_{-0.04}^{+0.04}$   &   $0.84_{-0.06}^{+0.06}$&   $0.56_{-0.06}^{+0.06}$ & $0.699_{-0.03}^{+0.03}$  & $0.70_{0.03}^{+0.03}$ &  $0.75_{-0.04}^{+0.04}$ \, (7.0$\sigma$)    \\
    
       \hline
      \textit{Planck-corr} & all &  DES+KiDS & $0.72_{-0.04}^{+0.04}$  &   $0.87_{-0.06}^{+0.06}$   &   
      $0.58_{-0.05}^{+0.05}$ &    $0.83_{-0.07}^{+0.07}$   &  $0.73_{-0.03}^{+0.03}$   & $0.79_{-0.03}^{+0.03}$ \, (6.8$\sigma$) \\
      &  all & HSC & $0.76_{-0.08}^{+0.08}$ &   $0.83_{-0.07}^{+0.08}$ &   
      $0.72_{-0.06}^{+0.06}$ &   $0.99_{-0.10}^{+0.10}$   &  $0.79_{-0.04}^{+0.04}$   & $0.84_{-0.05}^{+0.05}$   \, (3.5$\sigma$)\\

	\hline
	    \hline
     \textit{Lensing} & LS & DES+KiDS &$1.01_{-0.09}^{+0.09}$ &   $0.95_{-0.11}^{+0.11}$ &   $0.68_{-0.14}^{+0.12}$ &   $0.98_{-0.17}^{+0.16}$  & $0.93_{-0.06}^{+0.06}$  & $0.99_{-0.06}^{+0.06}$ \, (0.2$\sigma$)\\
     & SS & DES+KiDS &  $0.79_{-0.05}^{+0.04}$  &   $0.89_{-0.05}^{+0.05}$ &    $0.76_{-0.05}^{+0.05}$ &   $0.84_{-0.08}^{+0.09}$  &  $0.82_{-0.03}^{+0.03}$  &  $0.83_{-0.03}^{+0.03}$  \\
     
      \textit{Lensing-corr}& SS & DES+KiDS & $0.83_{-0.06}^{+0.06}$ &   $0.98_{-0.07}^{+0.07}$ &  $0.59_{-0.65}^{+0.72}$ &   $0.85_{-0.11}^{+0.10}$   &  $0.82_{-0.03}^{+0.03}$  & $0.89_{-0.04}^{+0.04}$ \, (2.8$\sigma$)\\

     	\hline
	\textit{Lensing-corr} & all &DES+KiDS & $0.84_{-0.05}^{+0.05}$  &   $1.00_{-0.07}^{+0.07}$  &  $0.67_{-0.06}^{+0.06}$  &   $0.97_{-0.08}^{+0.08}$   &  $0.85_{-0.03}^{+0.03}$  & $0.91_{-0.04}^{+0.04}$ \, (2.3$\sigma$)\\
		 & all & HSC & $0.88_{-0.09}^{+0.09}$  &   $0.96_{-0.09}^{+0.09}$  &   $0.85_{-0.07}^{+0.07}$  &   $1.16_{-0.12}^{+0.12}$  &  $0.93_{-0.04}^{+0.04}$ & $0.97_{-0.06}^{+0.06}$ \, (0.5$\sigma$)\\

     \hline
\end{tabular}
\caption{Results: 1D posterior constraints on the consistency scaling parameter, $A$, as shown in Figure~\ref{fig:alens} for each redshift bin (columns) and each scale range (rows): for all scales (all; $0.15<R<60h^{-1}{\rm Mpc}$), small scales (SS; $0.15<R<5.25h^{-1}{\rm Mpc}$) and large scales (LS;  $5.25<R<60h^{-1}{\rm Mpc}$), analysed in both the \textit{Planck} and \textit{Lensing} cosmological framework (upper and lower panels), for DES+KiDS (and HSC). In addition, we report the constraints for the cases where we account for baryonic effects and assembly bias, `corr'. For HSC we do not have the signal to noise at large scales to measure the consistency at an $R>5.25 h^{-1}\mathrm{Mpc}$. We additionally compute the inverse-variance weighted average of L1, L2 and C2, assuming the lens bins are independent, which will result is a stronger deviation from $A=1$ than is actually present in the data. As the large-scale check fails for bin C1, with a constraint on $A$ that deviates from 1, which we infer as evidence for systematics, we neglect this bin (denoted as an * in the table).}
\label{tab:Alens}
\end{table*}

First, as a consistency check, we revisit large scales where baryonic effects and assembly bias are negligible and fit the data including the inconsistency parameter, $A$. We expect that $A=1$ in $\Lambda$CDM on large scales in the absence of systematics, similar to the expectation for $X_{\rm lens}$ described in the combined linear-regime lensing and clustering analysis from \citet{y3-2x2RM, y3-3x2ptkp}. 
As before, we only have precise DES+KiDS measurements at these scales and do not include HSC. In the left-hand panel of Figure~\ref{fig:alens}, we show the 1D posteriors on the $A$ parameter for each lens redshift bin, for both a \textit{Planck} (red circles) and \textit{Lensing} cosmology (blue diamonds), and report the constraints in Table~\ref{tab:Alens}. These are consistent with $A=1$ for both cosmologies. This is in agreement with our expectations from the previous section, where the model provides a good fit to lensing and clustering. The exception is lens bin C1, where we find a substantial ($\sim3\sigma$) inconsistency. Note that the inclusion of the scaling parameter does improve the goodness-of-fit (now a p-value of 0.18) compared to that found in Section~\ref{sec:joint} when assuming a \textit{Planck} cosmology. 
As only one bin fails the consistency check, this may indicate the presence of unaccounted for systematics. 
For example, the result could be explained by selection effects in that particular lens sample that are not well modelled; it is unlikely that either using an alternative, but realistic $S_8$ value or accounting for astrophysical effects could resolve this result, as also found in the similar analysis of \citet{y3-2x2RM}. BOSS selection effects create four different lens samples, as illustrated in Figure~\ref{fig:nz_pervol}, which shows the redshift dependence of the comoving number density of galaxies. While L1, L2, and C2 are closer to flux-limited samples, for the C1 redshift range of $z=0.43-0.54$, this function is sharply increasing. To understand this fully would require further investigation into BOSS selection effects which are beyond the scope of this study. Given our findings, we consider C1 as an outlier result, and neglect it when computing a combined constraint on $A$ from the lens bins. However, as this analysis was not performed in a blind manner, in the interest of transparency, we include C1 in all figures and tables. For an overall constraint on $A$, we compute the inverse-variance weighted average of L1, L2, and C2, approximating the bins to be independent of each other. Note that this assumption does not account for any covariance between the lens bins; our combined estimate will result in an overestimation of the deviation from $A=1$. Using these three bins, we find that $A=0.99\pm0.06$ for the \textit{Lensing} cosmology, and  $A=0.86\pm0.06$ for the \textit{Planck} cosmology. These results support those from Section~\ref{sec:joint} that the large-scale data alone lacks the statistical power to significantly distinguish between these two cosmologies. 

Next, we consider scales less than $5.25h^{-1}$Mpc, with the results shown as the centre panel of Figure~\ref{fig:alens}.  Here, we find that $A$ significantly deviates from unity when assuming a \textit{Planck} cosmology for all lens bins. This is only partially resolved when assuming a \textit{Lensing} cosmology, such that a low-$S_8$ cosmology mitigates but cannot fully explain the differences between small-scale lensing and clustering. This finding is in agreement with previous work \citep{Leauthaud:2017aa, Lange2019,Lange2021}. However, on these scales astrophysical effects, which act to suppress the amplitude of the lensing signal such that $A<1$ (see Section~\ref{sec:uncertainties}), must be accounted for. 

\subsection{Accounting for small-scale model uncertainties}\label{sec:stackerror}

We consider the known systematics discussed in Section~\ref{sec:uncertainties} when quantifying the consistency of lensing and clustering on non-linear scales. Until we can sufficiently account for our uncertainty when modelling assembly bias and baryonic effects, 
as demonstrated in Figure~\ref{fig:eacherror}, the precision of analysis of galaxy--galaxy lensing and clustering measurements extended to non-linear scales in cosmological inference will be hindered by systematic errors. 

We include systematic corrections for both baryonic effects and assembly bias, modelled as
\begin{equation}
    \Delta\Sigma_{\rm obs}^{\rm corr}(R)=A_{\rm sys}(R) \Delta\Sigma_{\rm obs}(R)
\end{equation}
where $A_{\rm sys}(R)$ is defined as a combination of $A_{\rm bary}(R)+A_{\rm ab}(R)$, assuming the systematics are uncorrelated and additive. These are estimated as the fractional impact, $[\Delta\Sigma_{\rm fconc=0.84}-\Delta\Sigma_{\rm th}]/\Delta\Sigma_{\rm th}$ and $[\Delta\Sigma_{\rm ab}-\Delta\Sigma_{\rm th}]/\Delta\Sigma_{\rm th}$, indicated in Figure~\ref{fig:eacherror} as the red line (lower panel) and blue line (middle panel), respectively. These systematic corrections have an associated uncertainty, $\sigma^2_{\rm sys}(R)=\sigma^2_{\rm bary}(R)+\sigma^2_{\rm ab}(R)$, which are indicated as the red and blue shaded regions in Figure~\ref{fig:eacherror}, encompassing the null hypothesis of zero baryonic feedback and assembly bias. We include this as an additional systematic uncertainty on our measurement and our covariance assuming that the off-diagonal terms are impacted in the same way as the diagonal part of the covariance. We emphasise here that the predicted suppression of the lensing due to baryons is still uncertain, with a range of values in the literature from various constraints or simulation-based estimates. 
The prediction we have assumed here has significantly smaller impact than the estimate from the kSZ analysis \citep{amodeo21}, as discussed in Section~\ref{sec:uncertainties} and demonstrated in Figure~\ref{fig:eacherror}. Our associated error budget does span a large range of amplitudes, however it does not account for the difference in scale dependence between the various estimates. 

In the middle panel of Figure~\ref{fig:alens}, we show the best fit $A$ for KiDS+DES small-scale measurements and the impact when correcting for these small-scale systematics, with constraints on $A$ reported as `corrected' (filled circles), as before, for the two cosmologies. We find that the combined effect of these systematics is to suppress the predicted lensing signal at small scales, therefore reducing the amplitude $A$ to be less than one. However, we find that accounting for these effects with our current best estimates cannot resolve the small-scale detection of $A \ne 1$. We find a discrepancy with $A=0.75\pm0.04$ for the case of the \textit{Planck} cosmology and $A=0.89\pm0.04$ for the \textit{Lensing} case.

Finally, the right-hand panel of Figure~\ref{fig:alens} shows the posteriors for the full-scale measurements from both DESY3+KiDS1000 (filled green) and HSC (filled yellow), corrected for modelling systematics, for both the \textit{Planck} (red outer) and \textit{Lensing} (blue outer) cosmology; see also Table~\ref{tab:Alens}. The clustering and lensing measurements are presented in Appendix~\ref{app:fits}, along with the all-scale best-fit model for both cases of the \textit{Planck} and \textit{Lensing} cosmology when including the $A$ inconsistency parameter. For all bins, the consistency between the clustering and lensing measurements is improved in the \textit{Lensing} cosmology. In this cosmology and considering bins L1, L2 and C2, the measurements find $A=0.91\pm0.04$ for DES+KiDS and $A=0.97\pm0.06$ for HSC, consistent with $A=1$ within $\sim2\sigma$. For the \textit{Planck} cosmology, however, the combined result is found to be $A=0.79\pm0.03$ for  DES+KiDS and $A=0.84\pm0.05$ for HSC, or a $\sim3-7\sigma$ finding of $A \ne 1$.

Overall, with our theoretically--reasoned corrections for assembly bias and baryonic effects applied to the vanilla HOD model, we find that the lensing and clustering are consistent in a \textit{Lensing} cosmology. As the large-scale measurements of the two probes in Section~\ref{sec:joint} found similarly good fits to the data in the two cosmologies, at present, cosmological differences are driven by the small scales. Section~\ref{sec:uncertainties} demonstrated that there is uncertainty in the amplitude and extent of baryonic effects. Although our fiducial approach incorporates this uncertainty, it does not fully capture the scale-dependence of this effect that is indicated by some current kSZ measurements. Therefore, it is still possible and an open question as to whether the deviation from $A=1$ can be explained by both a \textit{Lensing} cosmology or a \textit{Planck} cosmology with a larger contribution from assembly bias and baryonic effects.
To be able to distinguish these two scenarios will require more precise measurements of the GGL signal at large scales and an improved understanding of the possible extent of baryonic effects and assembly bias. Section~\ref{sec:SZ} provides a next step to addressing that question.

\section{Cosmology \& baryonic effects}\label{sec:outlook}\label{sec:SZ}

\begin{figure}
\centering
\includegraphics[width=\columnwidth]{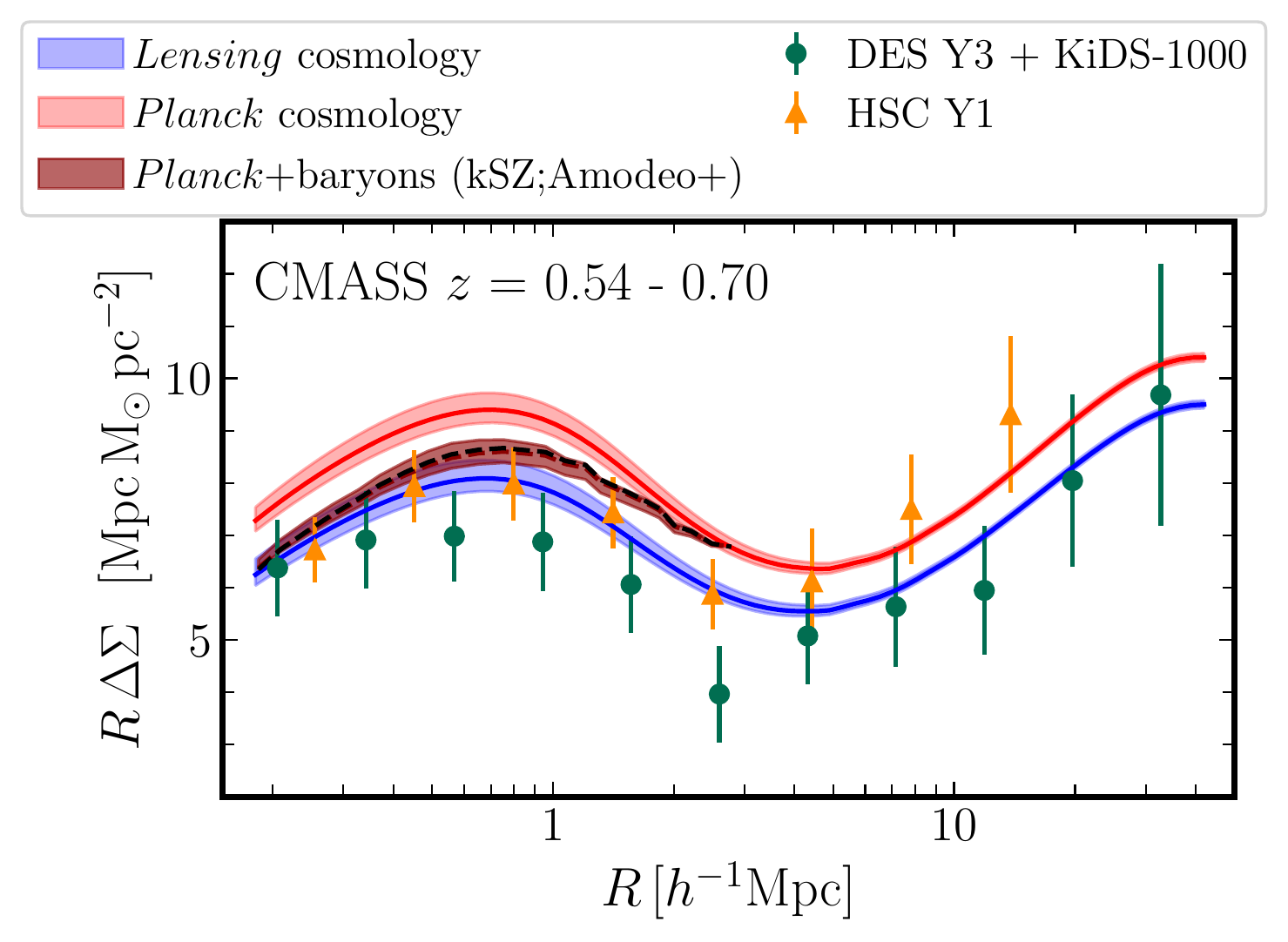}
\caption{The galaxy--galaxy lensing measurements for the C2 BOSS bin as measured by KiDS+DES (green) and HSC (yellow), alongside clustering-based HOD predictions for the signal derived at fixed \textit{Planck} (red) and \textit{Lensing} cosmology (blue). Note that this approach neglects the theoretically--reasoned model correction and uncertainties described in Section~\ref{sec:stackerror}. The shaded red and blue represent the 68\% confidence region of the prediction based on the posterior of an HOD fit to the clustering measurements from that bin. The dark red line (shaded region) represents the HOD model prediction made at \textit{Planck} cosmology, now including a correction for baryonic effects (although still neglecting uncertainties and assembly bias), based on the best fit density profile from kSZ measurements, as presented in \citet{amodeo21}. The baryon-corrected prediction is better able to reconcile the observations, and shows the degeneracy of the small-scale modelling corrections with the \textit{Lensing} cosmology prediction. This motivates future studies to investigate the impact of baryons on GGL and clustering studies.}
\label{fig:kSZ}
\end{figure}

There is no compelling evidence for a significant mismatch in the amplitude of clustering between galaxy auto-correlations and lensing observables on the easier-to-model linear scales; the largest inconsistencies are driven by small scales. As previously discussed, tension on non-linear scales could be the interaction of numerous hard-to-model physical effects of approximately comparable amplitude. Notably the impact of baryonic feedback on the mass and galaxy distributions in group-sized haloes at intermediate redshift is poorly understood, as evidenced by the scatter in the lower panel of Figure~\ref{fig:eacherror} and discussed in detail in Section~\ref{sec:uncertainties}. In our fiducial analysis we account for this effect  by varying $f_{\rm conc}$ ( Section~\ref{sec:stackerror}), but we note that this approach does not capture the scale dependence exhibited by the kSZ-based prediction for the impact of baryons presented in \citet{amodeo21}. In this section, we demonstrate how upcoming joint analyses with such data can shed light on the small-scale differences. 

We revisit an approach first presented in \citet{Leauthaud:2017aa}: we fit an HOD model to projected clustering measurements and use the clustering-based HOD parameters to predict the lensing signal, $\Delta\Sigma(R)$, which we then compare to the lensing measurements. Note that in this comparison, contrary to the previous section, baryonic effects and assembly bias and their associated uncertainties are not accounted for. 
The results are shown for the C2 case in Figure~\ref{fig:kSZ}:  the clustering-based lensing predictions and the lensing measurements for both cases --- assuming the \textit{Planck} cosmology (red) and the \textit{Lensing} cosmology (blue) -- with the shaded region indicating the model uncertainty derived from the statistical error on the $w_{\rm p}$ measurements. We compare to the DESY3+KiDS1000 (green) and HSC (yellow) measurements for each of the four lens samples. 
As detailed in Table~\ref{tab:Alens}, we find a scale dependence in the consistency between the prediction and measurement too. When isolating the large scales ($>5.25h^{-1}$Mpc) the difference between the \textit{Lensing} and \textit{Planck} predictions is negligible: with the current statistical power of lensing+BOSS data, the linear-scale measurements cannot sufficiently distinguish between the two cosmologies, with at most $\sim2\sigma$ differences. When considering all scales ($>0.15h^{-1}$Mpc), and without accounting for assembly bias and baryonic effects, the lensing prediction and model appear more consistent in the low-$S_8$ cosmology. 

Figure~\ref{fig:kSZ} includes an additional prediction for the lensing signal (dark red dashed line and shaded region). This represents our fiducial \textsc{Dark Emulator} model prediction made at a \textit{Planck} cosmology modified to correct for the impact of kSZ-estimated baryonic effects, drawn from \citet{amodeo21}. Note that \textit{Planck}+kSZ prediction in Figure~\ref{fig:kSZ} does not account for the impact of assembly bias, and the \textit{Lensing} cosmology (blue) and \textit{Planck} cosmology (red) predictions do not account for either baryonic effects or assembly bias. As is the case for the \textit{Lensing} cosmology prediction, we see that this kSZ baryon-correction to the \textit{Planck} prediction helps reconcile the difference with small-scale observations. That is, both the \textit{Lensing} and \textit{Planck} cosmologies provide a good fit to the large-scale data and the statistical power of the small-scale lensing measurements ($<5h^{-1}$Mpc) drive the preference for a \textit{Lensing} cosmology. However, in this case where kSZ measurements are available, we see evidence of a second viable solution to the previously-reported inconsistency between lensing and clustering: a \textit{Planck} cosmology with a kSZ-estimated correction for the extent of baryonic effects. 

These findings do have associated caveats which are detailed in Section 4 of \citet{amodeo21}, the most notable being that the model used in the kSZ analysis is normalised to the cosmic fraction at large radii and dark matter back reaction --- the effect of including baryons which alters the dark matter distribution due to the change in the gravitational potential --- is not included. Additionally the CMASS galaxies are weighted differently in \citet{amodeo21} to the GGL measurements discussed in this work, but we assume that the relative impact of baryonic effects does not vary substantially across the redshift range of CMASS. Given these caveats, we do not draw definitive conclusions, but qualitative ones. Nevertheless these results are compelling and motivate future studies that use SZ information to investigate the impact of baryons on lensing and clustering studies. Moreover, the comparison of these two probes can be used to place constraints on the extent and amplitude of the impact of baryonic effects. 

\vspace{0.5cm}
\section{Summary and Conclusions}\label{conclusions}

In this study we evaluate the consistency between lensing and clustering probes of large-scale structure based on measurements of projected
galaxy clustering from BOSS combined with overlapping galaxy--galaxy lensing
from three surveys: DES Y3, HSC, and KiDS-1000. We consider systematics in both the data and the modeling of these measurements. To assess the consistency, we perform a joint fit of the clustering and lensing measurements including a scale-independent multiplicative systematic parameter, $A$, which decouples the two signals and captures any inconsistency. Given additional uncertainties inherent in modelling the non-linear regime, we investigate the scale dependence of our results.
Our work builds upon previous efforts that have assessed the comparison between galaxy--galaxy lensing and projected clustering with BOSS \citep[e.g.][]{Leauthaud:2017aa, Lange2019, Wibking_2019,Yuan:2020,Lange2021,amodeo21,Yuan:2021}:  
\begin{itemize}
\item We update the galaxy--galaxy lensing measurements to use data from the three current state-of-the-art lensing surveys and include their calibration errors. Previous work has focused on CFHTLenS, CS82, and SDSS.
\item In the modelling, we use the {\sc Dark  Emulator}, which has demonstrated percent-level accuracy  in the transition regime, which is improved compared to analytic work \citep{DQHM,Mahony2022}. We note that \citet{Wibking_2019, Yuan:2020} also take an emulator approach. 
\item We simultaneously account for the impact of and uncertainty associated with lensing calibration systematics (including redshift calibration and blending), lens magnification, baryonic feedback, and assembly bias. Some of these effects have previously been considered individually.
\item We assess consistency by jointly fitting the lensing and clustering with an inconsistency parameter. 
\item In order to assess the scale dependence of the results, we repeat the joint fits limited to large scales, as GGL measurements are non-local and sensitive to matter at smaller scales (unlike clustering). 
\end{itemize}

We perform our joint lensing and clustering analysis in a $\Lambda$CDM framework at fixed cosmology and consider two cases: a \textit{Planck} cosmology and a low-$S_8$ \textit{Lensing} cosmology -- defined with a lower value of the $S_8$ parameter, drawn from \citet{Heymans2021}. Our primary findings are as follows:
\begin{itemize}
\item We update the \citet*{LWB} lensing data comparison based on their GGL measurements with BOSS. We find good agreement between DES-Y3, KiDS-1000, and HSC-Y1 data in all lens redshift bins and across all angular scales. Limited by the overlapping on-sky footprints of the surveys, we combine our measurements from KiDS and DES, but not HSC. 
\item Isolating large scales ($>$5.25$h^{-1}$Mpc) where the impact of baryonic effects and assembly bias are negligible, we perform a joint fit to the clustering and lensing  and find that both cosmologies provide an acceptable fit to the data (p-value$>0.01$). We find a mild preference for a low-$S_8$ \textit{Lensing} cosmology compared to the \textit{Planck} CMB cosmology. This preference is weaker than that deduced from cosmic shear surveys given the limited statistical power when considering only the linear regime. 
\item We revisit the fit to large scales ($>$5.25$h^{-1}$Mpc) including the inconsistency parameter, $A$, as a consistency check. In this linear regime, we expect that $A = 1$ in $\Lambda$CDM in the absence of systematics. We find consistency with $A=1$ in both cosmologies for bins L1, L2 and C2. C1 fails this consistency check in both cosmologies ($\sim3\sigma$), distinguishing this bin as an outlier. This may indicate unmodelled systematics in this bin, and thus for subsequent results, we compute a combined constraint on $A$ as the inverse-variance weighted average of L1, L2, and C2.
\item We estimate the impact of assembly bias and baryonic effects. For the former, we find that assembly bias can suppress the lensing signal at small scales by up to $\sim$15\%, but is negligible at scales $>$5.25$h^{-1}$Mpc. Estimates of baryon feedback also suppress the lensing profile, however the extent of the impact --- both the maximal amplitude and scale dependence --- depends on the simulation chosen or approach taken to estimate it. Our fiducial estimate of the impact of baryons suppresses the lensing signal at small scales by up to $\sim$10\% and is negligible above 1$h^{-1}$Mpc. We caution that the impact of these astrophysical effects on lensing measurements is still not sufficiently understood. 
\item We include a wider range of scales in the joint fit ($0.15<R<60h^{-1}$Mpc) and account for baryonic effects and assembly bias with a reasoned theoretical systematic uncertainty. In this case, we find $A=0.91\pm0.04$ using DESY3+KiDS1000 and $A=0.97\pm0.06$ using HSC and conclude that the measurements are consistent with $A=1$ when a low-$S_8$ \textit{Lensing} cosmology is assumed. Assuming a \textit{Planck} cosmology, on the other hand, we find $A$ deviates from 1, with $A=0.79\pm0.03$ (6.8$\sigma$) using DESY3+KiDS1000 and $A=0.84\pm0.05$ (3.5$\sigma$) using HSC. 
\item Qualitatively, we explore an alternative estimate of the impact of baryonic effects using kinematic Sunyaev--Zeldovich measurements from \citet{amodeo21} in the C2 bin, which exhibits a difference in scale-dependence of the effect compared to our fiducial approach. We find that we are better able to achieve consistency between lensing and clustering in a \textit{Planck} Universe with this assumption. Given current uncertainties in modelling these small-scale effects, this suggests that 
there may also be an astrophysical solution to the `Lensing is Low' problem and
it is important to continue to investigate both astrophysical and cosmological avenues.
\end{itemize}

The use of these highly non-linear scales incurs the challenge of estimating a theoretical error budget to account for astrophysical effects. Given the uncertainty on the extent and scale--dependence of small-scale baryonic effects and assembly bias, at present, cosmological inference deduced from an HOD analysis of these scales should be considered with caution. There is a pressing need to consider the implications of complex galaxy selections like those postulated to most strongly impact the C1 BOSS lens bin of this work, as well as to fully consider and more tightly bound the uncertainty range for HOD modelling systematics and astrophysical effects. Our work indicates how powerful these scales can be, if the possible impact of small-scale systematics can be robustly bounded.

The upcoming Dark Energy Spectroscopic Instrument \citep[DESI, ][]{DESI2013} survey will measure redshifts for millions of galaxies with a sky footprint roughly 14,000 deg$^2$. The cross-correlation of these spectroscopic survey redshifts and those from the Subaru Prime Focus Spectrograph \citep[PFS, ][]{PFS} with the imaging surveys used in this analysis (KiDS-1000, DES Year 3, and HSC) will allow for scrutiny of both the consistency and the joint analysis with measurements of the clustering in unprecedented detail. Using BOSS, this work is a stepping-stone towards a unified galaxy-galaxy lensing analysis with these imaging surveys applied to DESI data in the future.
Furthermore, our findings motivate additional work that includes modelling beyond a simple HOD. In the future, when working with lens samples with complicated colour selections, the analysis can be limited to volume-limited samples (e.g. the luminosity-limited sample used in \citet{HSC2x2}). In addition, the lensing data can be more conservatively selected to limit the source--lens redshift overlap. 

There are a number of steps that can be taken to enable future robust cosmological constraints that incorporate measurements from small scales.
One can consider joint analyses with RSD as opposed to projected clustering in order to jointly model assembly bias \citep[e.g.][]{Yuan:2021}. A promising avenue to understand the impact of baryonic effects is through cross-correlation and joint analyses with the measurements of the Sunyaev--Zeldovich effect, which can independently constrain the distribution of baryons \citep[e.g.][]{troester2021, gatti2021,pandey2021}. An approximation of this, using the kinematic Sunyaev--Zeldovich and galaxy--galaxy lensing measurements for the same lenses was demonstrated in \citet{amodeo21}, and work is currently underway with Advanced ACT to instead, jointly analyse the probes. Future CMB  \citep{SO,CMB-S4} and lensing observations from Rubin and \textit{Euclid} analysed with DESI and PFS spectroscopy will further this effort with higher sensitivity measurements that allow for probing baryonic effects as a function of mass, redshift, and galaxy properties.

\section*{Acknowledgements}
We thank Stefania Amodeo and Nicholas Battaglia for sharing the baryon predictions from \citet{amodeo21} shown in Figure~\ref{fig:kSZ} and Hong Guo for sharing the BOSS clustering measurements. 
We thank George Efstathiou, Johannes Lange, and Alexie Leauthaud for many useful discussions during the preparation of this manuscript and comments that improved the draft. We are also grateful for the feedback from the KiDS, DES and HSC teams on this manuscript, and for the enjoyable cross-survey collaborative experience.

This research manuscript made use of Astropy \citep{astropy:2013,astropy:2018} and Matplotlib \citep{matplotlib}, and has been prepared using NASA's Astrophysics Data System Bibliographic Services.
The authorship list reflects the two lead authors (AA, NR). Authors HM, CH, MW, JD, SY, RHW, TNV, SB, AD, SM, AR, HH contributed ideas and components of the paper. Authors who made infrastructure contributions to the DES Year 3 and KiDS-1000 data and the Dark Emulator form the first alphabetical group, and the second is for DES and KiDS builders. 

AA acknowledges financial support from the award of a Kavli Institute Fellowship at KIPAC and at KICC. HM and TN is supported by JSPS KAKENHI Grant Number 19H00677, by Japan Science and Technology Agency (JST) CREST JPMHCR1414, and by JST AIP Acceleration Research Grant Number JP20317829. HM is supported by JSPS KAKENHI Grant Numbers 20H01932 and 21H05456, by JSPS Core-to-Core Program Grant Numbers JPJSCCA20200002 and JPJSCCA20210003. CH, MA and TT acknowledge support from the European Research Council under grant number 647112. CH also acknowledges support from the Max Planck Society and the Alexander von Humboldt Foundation in the framework of the Max Planck-Humboldt Research Award endowed by the Federal Ministry of Education and Research. 
TN is supported in part by MEXT/JSPS KAKENHI Grant Numbers JP20H05861 and JP21H01081. MB is supported by the Polish National Science Center through grants no. 2020/38/E/ST9/00395, 2018/30/E/ST9/00698, 2018/31/G/ST9/03388 and 2020/39/B/ST9/03494, and by the Polish Ministry of Science and Higher Education through grant DIR/WK/2018/12. JTAdJ is supported by the Netherlands Organisation for Scientific Research (NWO) through grant 621.016.402. HHi is supported by a Heisenberg grant of the Deutsche Forschungsgemeinschaft (Hi 1495/5-1), and with AD and AW acknowledges support from an ERC Consolidator Grant (No. 770935). HHo acknowledges support from Vici grant 639.043.512, financed by the Netherlands Organisation for Scientific Research (NWO). HYS acknowledges the support from CMS-CSST-2021-A01 and CMS-CSST-2021-B01, NSFC of China under grant 11973070, the Shanghai Committee of Science and Technology grant No.19ZR1466600 and Key Research Program of Frontier Sciences, CAS, Grant No. ZDBS-LY-7013. TT acknowledges support from the Leverhulme Trust.

This KiDS data are based on observations made with ESO Telescopes at the La Silla Paranal Observatory under programme IDs 177.A-3016, 177.A-3017, 177.A-3018 and 179.A-2004, and on data products produced by the KiDS consortium. The KiDS production team acknowledges support from: Deutsche Forschungsgemeinschaft, ERC, NOVA and NWO-M grants; Target; the University of Padova, and the University Federico II (Naples).

Funding for the DES Projects has been provided by the U.S. Department of Energy, the U.S. National Science Foundation, the Ministry of Science and Education of Spain, 
the Science and Technology Facilities Council of the United Kingdom, the Higher Education Funding Council for England, the National Center for Supercomputing 
Applications at the University of Illinois at Urbana-Champaign, the Kavli Institute of Cosmological Physics at the University of Chicago, 
the Center for Cosmology and Astro-Particle Physics at the Ohio State University,
the Mitchell Institute for Fundamental Physics and Astronomy at Texas A\&M University, Financiadora de Estudos e Projetos, 
Funda{\c c}{\~a}o Carlos Chagas Filho de Amparo {\`a} Pesquisa do Estado do Rio de Janeiro, Conselho Nacional de Desenvolvimento Cient{\'i}fico e Tecnol{\'o}gico and 
the Minist{\'e}rio da Ci{\^e}ncia, Tecnologia e Inova{\c c}{\~a}o, the Deutsche Forschungsgemeinschaft and the Collaborating Institutions in the Dark Energy Survey. 

The Collaborating Institutions are Argonne National Laboratory, the University of California at Santa Cruz, the University of Cambridge, Centro de Investigaciones Energ{\'e}ticas, 
Medioambientales y Tecnol{\'o}gicas-Madrid, the University of Chicago, University College London, the DES-Brazil Consortium, the University of Edinburgh, 
the Eidgen{\"o}ssische Technische Hochschule (ETH) Z{\"u}rich, 
Fermi National Accelerator Laboratory, the University of Illinois at Urbana-Champaign, the Institut de Ci{\`e}ncies de l'Espai (IEEC/CSIC), 
the Institut de F{\'i}sica d'Altes Energies, Lawrence Berkeley National Laboratory, the Ludwig-Maximilians Universit{\"a}t M{\"u}nchen and the associated Excellence Cluster Universe, 
the University of Michigan, NFS's NOIRLab, the University of Nottingham, The Ohio State University, the University of Pennsylvania, the University of Portsmouth, 
SLAC National Accelerator Laboratory, Stanford University, the University of Sussex, Texas A\&M University, and the OzDES Membership Consortium.

Based in part on observations at Cerro Tololo Inter-American Observatory at NSF's NOIRLab (NOIRLab Prop. ID 2012B-0001; PI: J. Frieman), which is managed by the Association of Universities for Research in Astronomy (AURA) under a cooperative agreement with the National Science Foundation.

The DES data management system is supported by the National Science Foundation under Grant Numbers AST-1138766 and AST-1536171.
The DES participants from Spanish institutions are partially supported by MICINN under grants ESP2017-89838, PGC2018-094773, PGC2018-102021, SEV-2016-0588, SEV-2016-0597, and MDM-2015-0509, some of which include ERDF funds from the European Union. IFAE is partially funded by the CERCA program of the Generalitat de Catalunya.
Research leading to these results has received funding from the European Research
Council under the European Union's Seventh Framework Program (FP7/2007-2013) including ERC grant agreements 240672, 291329, and 306478.
We  acknowledge support from the Brazilian Instituto Nacional de Ci\^encia
e Tecnologia (INCT) do e-Universo (CNPq grant 465376/2014-2).

This manuscript has been authored by Fermi Research Alliance, LLC under Contract No. DE-AC02-07CH11359 with the U.S. Department of Energy, Office of Science, Office of High Energy Physics.

\section*{Data availability}
The data underlying this article cannot be shared publicly due to collaboration embargoes. The data will be shared on reasonable request to the corresponding author.

\bibliographystyle{mnras}
\bibliography{bib.bib}
  
\FloatBarrier
\appendix
\section{Lensing without borders II}\label{sec:lwb}

\begin{figure*}
\centering
\includegraphics[width=\textwidth]{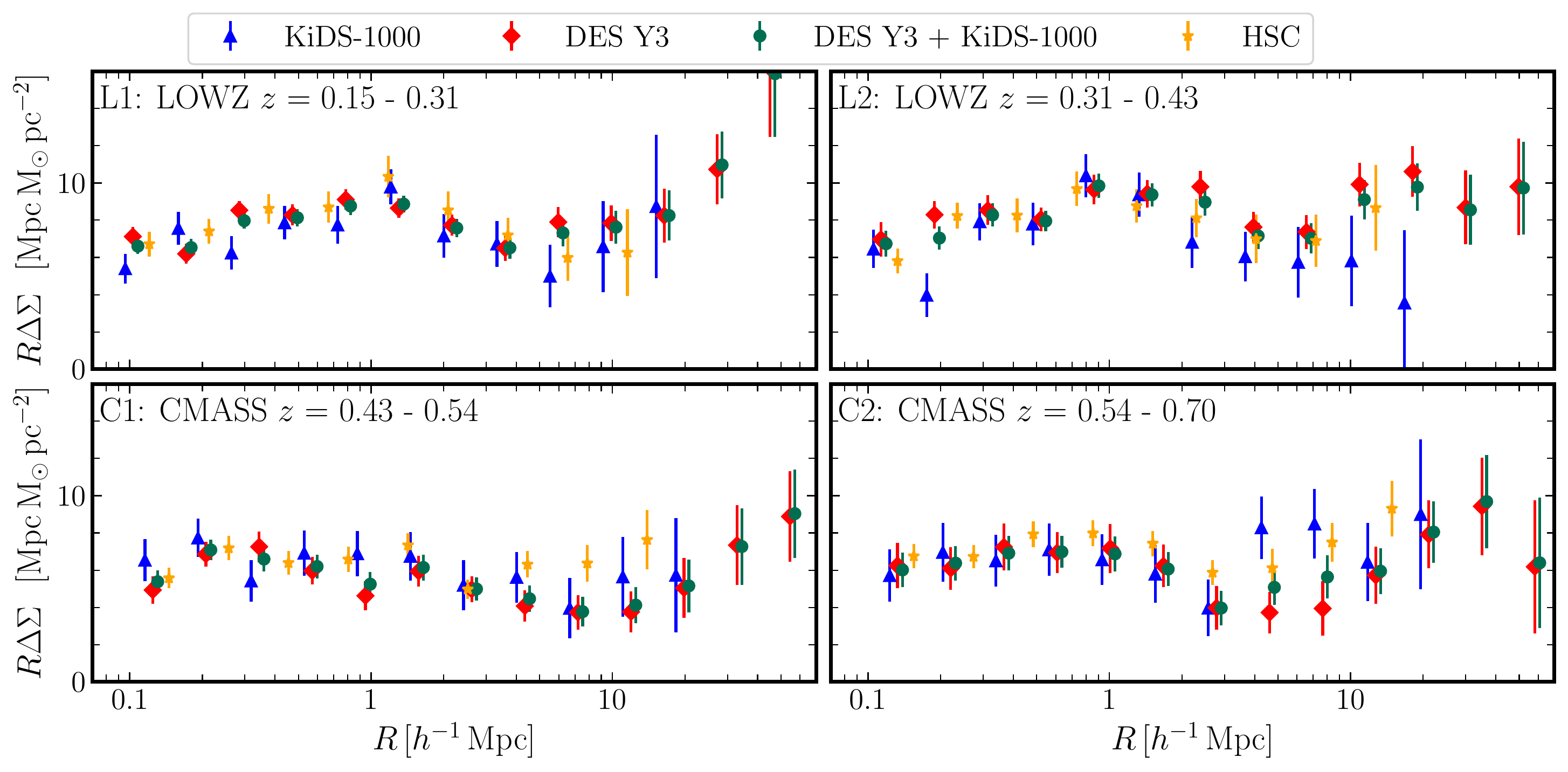}
\caption{The galaxy--galaxy lensing measurements, $\Delta\Sigma(R)$ around the four BOSS lens samples from HSC (yellow), KiDS-1000 (blue) and DES-Y3 (red). The green data points represent the inverse-variance weighted combined measurement of the latter two. The error bars combine both the statistical and systematic contributions to the uncertainty.}
\label{fig:ds}
\end{figure*}

In this analysis we extend the work of Lensing without borders I \citep*{LWB} by considering two new weak lensing data sets: KiDS-1000 and DES-Y3 presented in Section~\ref{sec:data}. The measurements from each of these are shown in Figure~\ref{fig:ds}.

\begin{table}
\centering
\begin{tabular}{cccccc}
	\hline
	Data & Stat & L1 & L2 & C1 & C2 \\
	\hline
	KiDS vs. DES & $\chi^2$/DoF  & 1.03 & 1.41 & 0.70 & 0.86 \\
        & p-value  & 0.49 & 0.20 & 0.82 & 0.66 \\
    KiDS+DES vs. HSC & $\chi^2$/DoF  & 0.28 & 0.76 & 1.61 & 0.74 \\
        & p-value  & 0.98 & 0.73 & 0.17 & 0.75 \\
	\hline
	
	 	\hline
\end{tabular}
\caption{We confirm consistency between our KiDS and DES measurements by estimating the $\chi^2$/DoF of the difference between the two measurements compared to null. From this we compute the corresponding p-value and find that for all bins this is greater than 0.2. A p-value less than 0.01 equates to a 99\% confidence in rejection of consistency between the data, assuming Gaussian statistics, and therefore the estimated p-value are not concerning. Since KiDS and DES have independent areas in common with BOSS we can combine them. This is not the case for HSC which overlaps with both the KiDS and DES footprint. We therefore check consistency between HSC and the combined KiDS+DES measurements with the same method and find consistency. When computing the covariance for each measurement we additionally include the systematic error.}
\label{tab:LWB}
\end{table}

The difference between the previous data from KiDS, KV450, and that presented in this work, KiDS-1000, is not merely an increase in area, but also reflects the development in methodology when analysing the lensing data. In \citet*{LWB}, which used the earlier KV-450 data, the redshift distributions were derived applying the weighted direct calibration (DIR) method \citep{Hildebrandt_2020}. This method was shown to produce well-calibrated redshift distributions for galaxies with photometric redshifts in the range $(0.1 < z_{\mathrm{B}} < 1.2)$. Here, following the recent cosmic shear analyses \citep{asgari20}, which prioritised the accuracy of the calibrated redshift distributions over the precision, we utilise the self-organising map approach (briefly described in Section~\ref{sec:datakids} and detailed in \citealt{wright20}). Due to the more stringent selection criteria applied to source galaxies, the galaxy number density is reduced in comparison to KV450, such that the estimate of the error on our new measurements with KiDS 1000 only improves by $\sim 20$\%, even though the area used has almost doubled. Conversely, the extra care taken estimating the redshift distribution leads to a reduction in the systematic error budget for both photometric redshift and multiplicative bias. 

Similarly, the DES lensing data calibration methodology has significantly changed between DES Year 1 (Y1) and Y3 \citep{y3-cosmicshear1}. First, the data processing included improved modelling of the Point Spread Function \citep{y3-piff}, and of the photometric calibration \citep{y3-gold}. Second, the significant improvements in the realism of the image simulations for Y3, presented in \citet{y3-imagesims}, as well as a more sophisticated understanding and modeling of the effects of blending that explicitly accounts for the impact of blending and detection biases as a function of redshift, including modifications to the effective $n(z)$. These Y3 multiplicative corrections result in a higher-amplitude lensing profile relative to those from DES Y1. Finally, a new framework for the redshift calibration was developed for the Y3 analysis \citep{y3-sompz}, built upon a self-organising map method. It employed a wide--deep survey strategy, using the DES Deep Fields \citep{y3-deepfields}, a combination of optical and near-infrared multi-band, deep photometry over a smaller area, to better characterise the colour-redshift relation, successfully reducing both the statistical and systematic uncertainty in redshift calibration. In addition, it accounts for biases that might arise due to the redshift sample's photometric outliers or spectroscopic incompleteness, and it characterises the impact of various sources of uncertainty on the full shape of the distribution \citep{y3-sompz}. 

In Section~\ref{sec:deskids}, we determine KiDS and DES measurements to be consistent and combine their measurements. Here we assess the consistency of the combined result with HSC, but cannot combine all three signals due to overlapping footprints on the sky. Comparing the DESY3$+$KiDS1000 measurements with the HSC measurements, we find the difference in signals compared to null has a p-value of $(0.98,0.73,0.17,0.75)$ and a reduced chi-squared value of $(0.28,0.76,1.61,0.74)$ for each redshift bin\footnote{The DoF are computed using mocks.}. The corresponding $\sigma$ values are $(0.02,0.35,1.38,0.32)$. By the same criteria used to confirm consistency between KiDS-1000 and DES Y3, given the p-value is always larger than 0.17, we conclude that these two GGL measurements are statistically consistent. This does not account for any correlation between the two data sets but is sufficient since we do not proceed with combining them.

\section{Magnification bias}\label{app:mag}

\begin{figure}
    \subfloat[]{\label{fig:dcellmag_dk}\includegraphics[width=1.0\columnwidth]{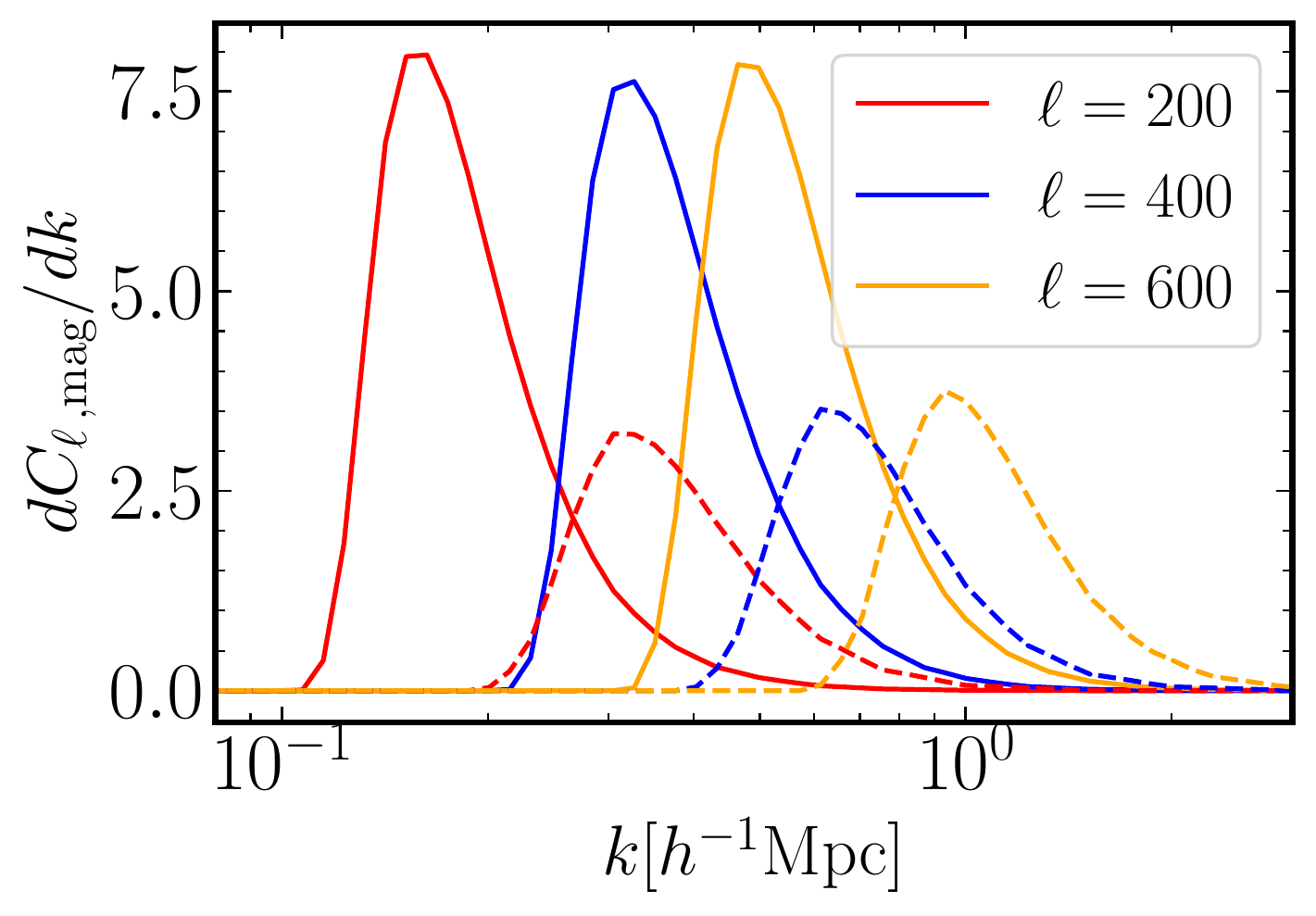}} \\
    \subfloat[]{\label{fig:mag}\includegraphics[width=1.0\columnwidth]{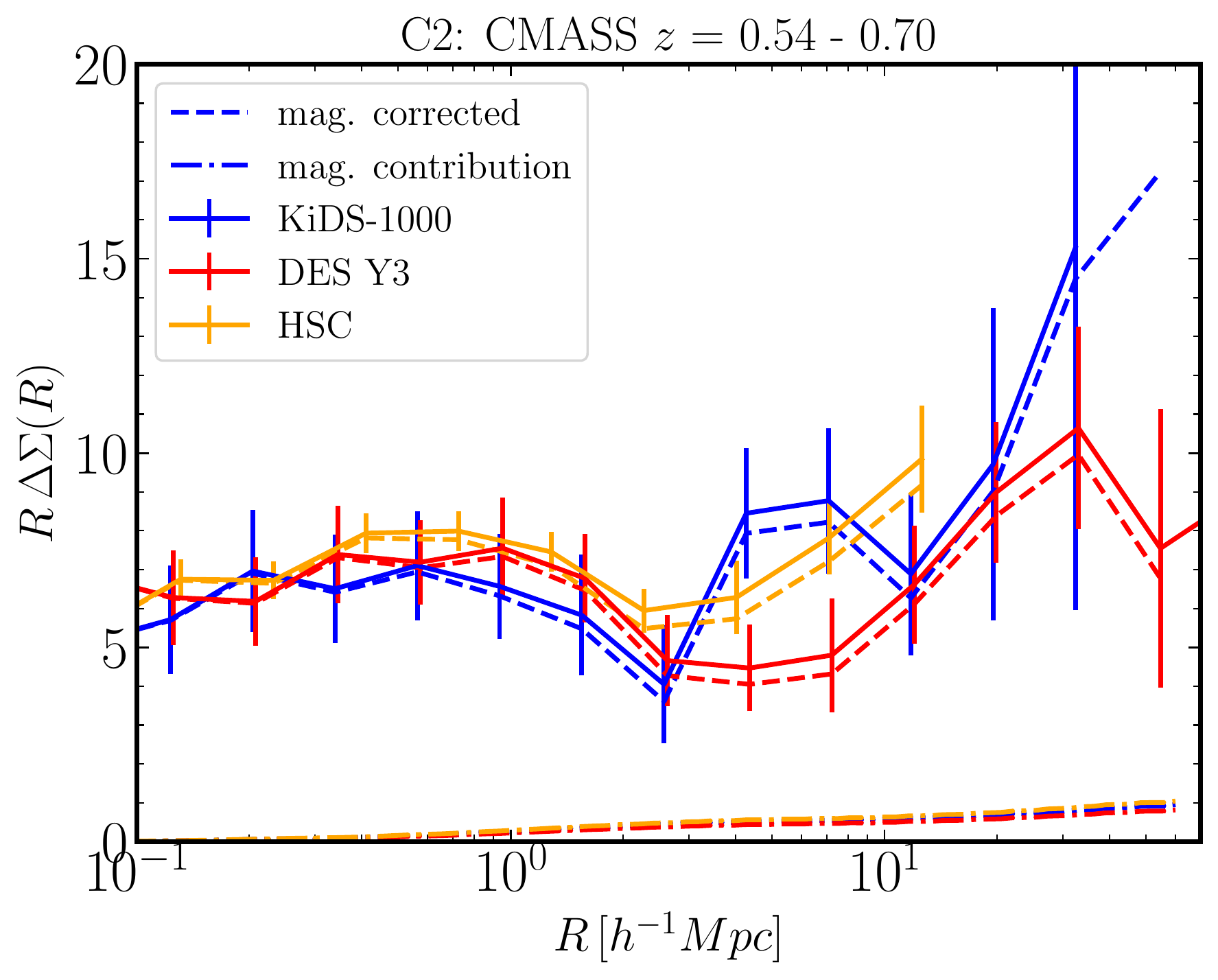}}
    \caption{Magnification correction. \ref{fig:dcellmag_dk}: Fractional contribution to the magnification angular power spectrum as a function of $k$ for $\ell = \{200, 400, 600\}$. Solid lines are for the C2 (CMASS $z=0.54-0.7$) lens sample, dashed are for L1, and both use KiDS-1000 fourth source bin. Even for low $\ell$, the magnification term receives significant contributions from high $k$, thus making it difficult to make angular scale cuts that localize the galaxy--galaxy lensing signal to large physical scales. \ref{fig:mag}: The impact of magnification on $\Delta\Sigma(R)$  for the C2 (CMASS $z=0.54-0.7$) lens sample, for KiDS (blue), DES (red) and HSC (yellow) signals. For each survey, the solid lines are the measurements, the dashed lines indicate the magnification-corrected measurements, subtracting the magnification contribution, which is shown in the dot-dashed line.}
    \label{mag}
\end{figure}

In Figure~\ref{fig:dcellmag_dk} we show the fractional contribution to $C_{\ell, {\rm mag}}$, defined in equation~\ref{eqn:mag}, as a function of $k$ for a few different values of $\ell$. It is apparent that even for relatively small $\ell$, the magnification contribution to $C_{\ell,\kappa g}$ receives significant power from high $k$, where the matter power spectrum is less well understood.  

As shown in Figure~\ref{fig:mag}, for the highest redshift lens sample considered in this work, which has the largest magnification contribution, the correction is less than a $10\%$ effect to the total galaxy--galaxy lensing signal. While the uncertainty correcting for this effect is sub-dominant for this analysis, it is important for future analyses to incorporate a model for the high-$k$ behavior of $P_{\rm m}$, even those that limit themselves to large angular scales.

\section{HOD modelling}
\label{app:model}
In this Appendix, we compare our fiducial {\sc Dark Emulator} model to one that includes incompleteness parameters, as well as a simulation-based model. Figure~\ref{fig:hod} shows the number of galaxies as a function of halo mass, i.e., the HOD model described in  Section~\ref{sec:ghc}, derived from the chains under the \textit{Lensing} cosmology. 

\subsection{Incompleteness}
\begin{figure*}
\centering
\begin{tabular}{@{}c@{}}
    \includegraphics[width=1.0\textwidth]{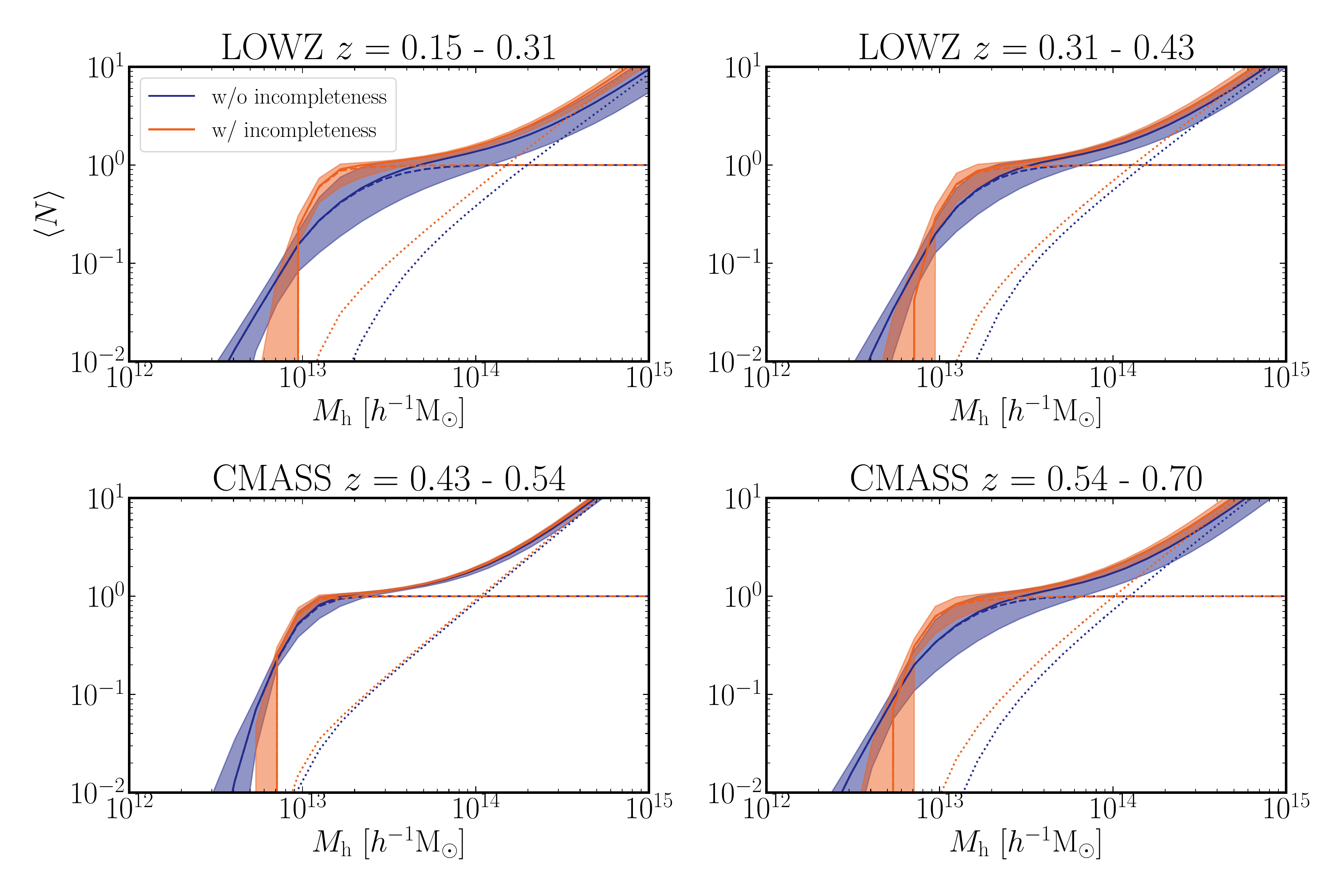}
 \end{tabular}
\caption{\label{fig:hod}The lines show the median of the HOD prediction for the expectation number of central (dashed), satellite (dotted) and the total (solid) number of galaxies as a function of the mass of the dark matter halo inside of which they reside for the clustering measurements presented in Section~\ref{sec:measurements}, derived from chains under the \textit{Lensing} cosmology. The orange and blue curves show the clustering fit with and without incompleteness in the model, respectively. The shaded regions show 68\% CL.}
\end{figure*}

In Figure~\ref{fig:hod}, the HODs of our fiducial setup adding the incompleteness model differ from that without incompleteness especially at the low mass end. However,
the clustering signals with and without the incompleteness do not show any significant difference, which means that the incompleteness model implemented in {\sc Dark Emulator} cannot reduce the difference between the data points and the model at large scales. The $\chi^2$ does not improve with the addition of the two incompleteness parameters but the effective degrees of freedom reduces so that the reduced-$\chi^2$ is degraded. This does not immediately mean the incompleteness model is not sufficient to model the incompleteness in the data, since we do not know if the discrepancy is from incompleteness or systematics in the measurement. However, this implies that careful modeling of incompleteness, e.g., based on simulations, will become important in the future cosmology experiments. We use the full BOSS sample in this paper, and it may have a complicated incompleteness as a function of colour and magnitudes \citep{Saito:2016}. This would make it difficult to model the incompleteness using the simple incompleteness model. Based on this investigation we choose not to include this incompleteness prescription in our fiducial analysis.

\subsection{Model variance}\label{HODimpl}

As a robustness test, we compare our fiducial model choice, {\sc Dark Emulator}, to an alternative approach used in previous literature: a simulation-based model, based upon \textsc{AbacusSummit}, used in estimating the assembly bias uncertainty \citep{Yuan:2020}. As these approaches incur several differences, we merely use the observed difference in their predictions as a gauge of model variance. A simulation-based approach \citep[see e.g.][]{Reid:2014, Leauthaud:2017aa, Saito:2016, McClintock2019} removes reliance on analytical fitting functions for scale-dependent halo bias and halo exclusion and is better able to accurately capture the properties of spectroscopic lens samples with selection effects that lead to redshift-dependent number densities and limit the ability of a single, redshift-independent HOD to accurately capture the properties of BOSS galaxies \citep{Saito:2016, Rodriguez-Torres:2015aa}. The development of emulators provide a means of delivering accurate predictions for multiple cosmological models \citep{Heitmann:2010, Wibking_2019, derose2019, DQ}. We use the same fit to the BOSS clustering measurements to predict the GGL signal, within the same cosmological framework. 
 
The simulation-based galaxy catalogs are generated on \textsc{AbacusSummit} \citep{Maksimova:2021}, which is a set of large, high-accuracy cosmological N-body simulations using the \textsc{Abacus} N-body code \citep{Garrison:2019, Garrison:2021b}, designed to meet the Cosmological Simulation Requirements of the Dark Energy Spectroscopic Instrument (DESI) survey \citep{DESI2013}. \textsc{AbacusSummit} consists of over 150 simulations, containing approximately 60 trillion particles at 97 different cosmologies. A typical base simulation box contains $6912^3$ particles within a $(2h^{-1}$Gpc$)^3$ volume, which yields a particle mass of $2.1 \times 10^9 h^{-1}M_\odot$. \footnote{For more details, see \url{https://abacussummit.readthedocs.io/en/latest/abacussummit.html}}. For this analysis, we primarily use the $z=0.3$ (LOWZ) and $z=0.5$ (CMASS) slices of the \verb+AbacusSummit_base_c000_ph000+ box, which adopts the \citet{Planck2018} $\Lambda$CDM cosmology and is $(2h^{-1}$Gpc$)^3$ in volume. The haloes are identified using the {\sc CompaSO} halo finder, which is an on-the-fly group finder specifically designed for the \textsc{AbacusSummit} simulations \citep{Hadzhiyska:2021}. A post-processing `cleaning' procedure leverages the halo merger trees to `re-merge' a subset of haloes \citep{Bose:2021}. This is done both to remove over-deblended haloes in the spherical overdensity finder, and to intentionally merge physically associated haloes that have merged and then physically separated. Finally, the galaxies are assigned onto haloes using the \textsc{AbacusHOD} code, which is a particle-based implementation of a generalized HOD model \citep{Yuan:2020, Yuan:2021b}. The HOD parameters are optimized to match the Dark Emulator's best fit to the projected clustering in each bin, and the best-fit HOD is used to predict the lensing signal.

Considering scales above 0.5$h^{-1}$Mpc, we find that predictions from the {\sc Dark Emulator} and \textsc{AbacusSummit} are consistent within 10\%. The differences found between the two approaches could be accounted for in the following ways; (i) redshift: the \textsc{AbacusSummit} predictions for L1 and L2 (C1 and C2) are both made using the simulation at redshift slice 0.3 (0.5), rather than tailoring to the individual redshift bins (ii) simulation differences such as force resolution and particle mass; it is possible that differences in resolution and force softening could lead to a more concentrated core profile in the haloes and therefore an excess in the cumulative lensing signal, illustrated in Figure~13 of \citet{Yuan:2021b};
(iii)  halo finder: the {\sc Dark Emulator} defines halo mass with the spherical overdensity ($M_{\rm 200m}$), \textsc{Abacus} simulations use a custom-developed spherical overdensity-based halo finder called {\sc CompaSO} \citep[see App C2 in][for more details]{Yuan:2021b}. 

While the difference between the two approaches is larger than each model uncertainty, 
we do not include an additional factor to account for this variance in our small-scale uncertainty, but note that the difference between the predictions warrants future work and a more detailed comparison, matching and varying each component listed above.

\subsection{Mis-centring}
\label{sec:miscentring}
We consider the impact of mis-centring of the lens galaxies \citep{Hoshino:2015} by accounting for the possibility that there is a fraction of central galaxies, $p_{\rm off}$, that are offset from the centre of their haloes \citep[see e.g.][]{Skibba:2011aa}.  As a simple model to estimate the possible impact of mis-centring on our observables we take the normalised radial profile of the mis-centered galaxies, 
with respect to the true halo centre, to be a Gaussian.  We set the Gaussian width to be a multiple, $R_{\rm off}$, of 
the spherical halo boundary radius within which the mean mass density is 200 times $\bar{\rho}_{\rm m}$, $r_{\rm 200m}$.
These two additional parameters enter our halo model through the central halo term in equation \ref{eqn:pgm}, modifying it as 
\begin{multline}
    \frac{\langle N_{\rm c} | M_{\rm h} \rangle}{n_{\rm gal}(z)} \rightarrow \\
    \quad\quad\quad \frac{\langle N_{\rm c} | M_{\rm h} \rangle}{n_{\rm gal}(z)} \left( 1 - p_{\rm off} + p_{\rm off}\, \exp \left[-\frac{1}{2} k^2 (r_{\rm 200m} R_{\rm off})^2 \right]\right) \, .
\end{multline}
To investigate the impact of mis-centring, we include the $p_{\rm off}$ and $R_{\rm off}$ parameters in our HOD fit to the clustering, using uniform priors $[0,1]$ as well as more informative priors derived from the posteriors reported in \citet{More:2015} of $R_{\rm off}= \mathcal{N}(0.5, 0.3)$ and $p_{\rm off}=\mathcal{N}(0.35,0.2)$. The galaxy clustering is insensitive to the values of the mis-centring parameters, recovering similar HOD parameters in the cases considered.  In contrast, the GGL signal on small scales is highly sensitive to this effect. 
We conclude that the clustering alone can not constrain the mis-centring parameters at the scales we consider and we do not include it in our fiducial model, nor have a well-informed uncertainty bound to report, such that it is not included in Section~\ref{sec:stackerror}. We include it here as a potential systematic for completeness. Additionally the baryon feedback error budget is sufficiently conservative to account for any mis-centering effects.

\section{Joint fits with $A$}
\label{app:fits}
\begin{figure*}
\centering
\includegraphics[width=\textwidth]{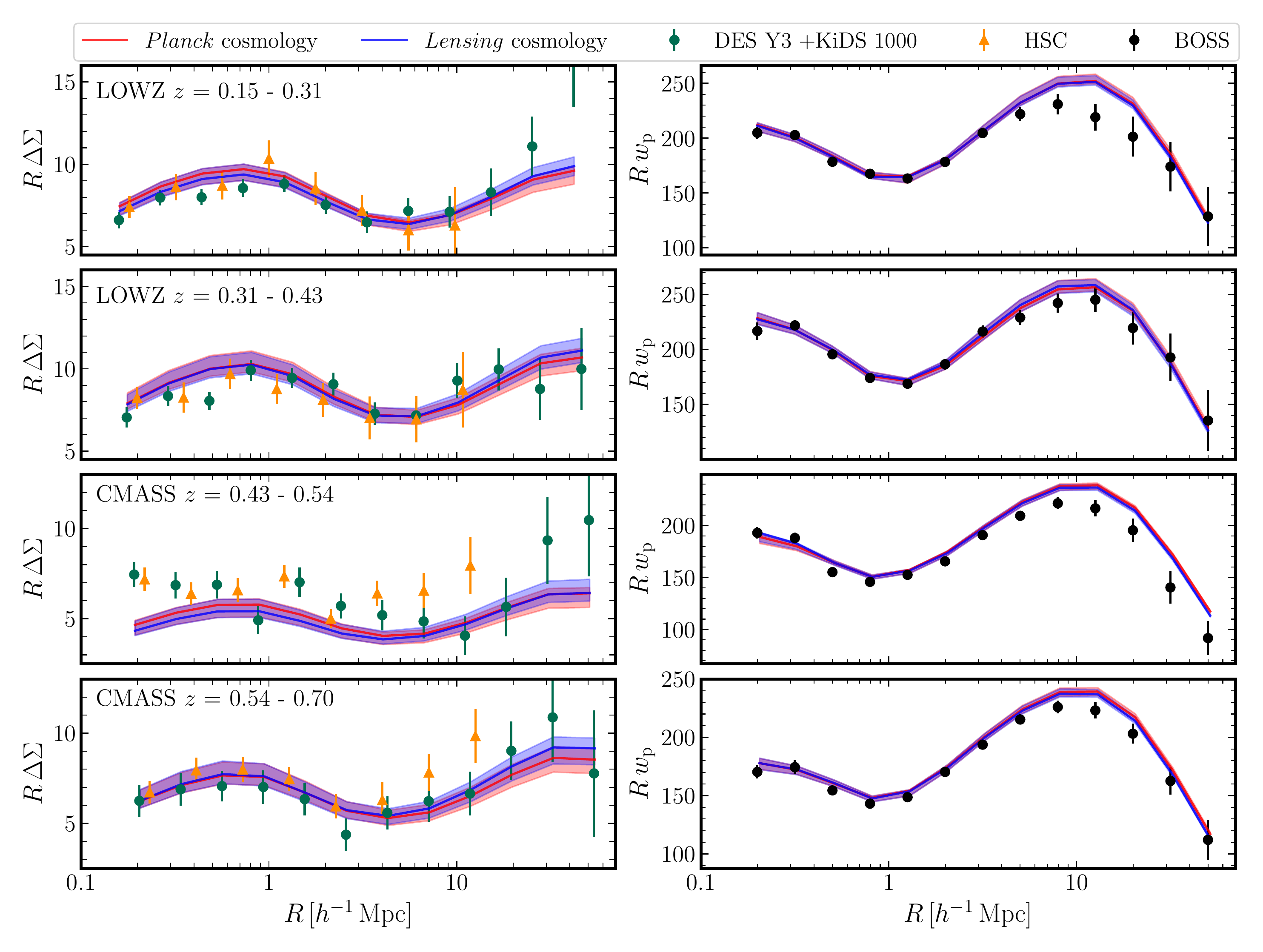}
\caption{Joint fits to the BOSS clustering measurements (black data points, right panels), $w_{\rm p}$, and lensing profiles, $\Delta\Sigma(R)$, measured by DES Y3 + KiDS-1000 (green, left panels), computed using the {\sc Dark Emulator} and including an inconsistency scaling parameter, $A$, for each of the four lens samples. We also show the HSC measurements (orange), but not the fits for this case. The fits are performed at a fixed cosmology, using parameters from \citet{Planck2018} (red), as well as a \textit{Lensing} cosmology (blue), using a lower value of $S_8$ \citep{Heymans2021}.  
The corresponding values of $A$ for each lens bin, which are reported in Table~\ref{tab:Alens}, are $A = [0.72\pm0.04, 0.87\pm0.06, 0.58\pm0.05, 0.83\pm0.07]$ assuming a \textit{Planck} cosmology and $A = [0.84\pm0.05, 1.00\pm0.07, 0.67\pm0.06, 0.97\pm0.08]$ assuming the \textit{Lensing} cosmology for the DES Y3 + KiDS-1000 data, and for the 
HSC data $A=[0.76\pm0.08, 0.83\pm0.08, 0.72\pm0.06, 0.99\pm0.10]$ and $A=[0.88\pm0.09, 0.96\pm0.09, 0.85\pm0.07, 1.16\pm0.12]$ assuming a \textit{Planck} and \textit{Lensing} cosmology respectively.
The lines indicate the best-fit model, and the corresponding shaded regions represent the 68\% confidence level derived from the posterior of each of the fits. For the clustering, the fit is largely independent to this change in cosmological parameters, while the difference is seen for the lensing. When limited to the large scales, the \textit{Lensing} cosmology is preferred, although the difference between the fits is not significant.}
\label{fig:cosmo}
\end{figure*}

In this appendix we present our measurements for clustering and lensing across the full range of scales ($0.15h^{-1}{\rm Mpc}< R<60h^{-1}{\rm Mpc}$) and the corresponding best fit model from the joint fits discussed in Section~\ref{sec:Alens}. Figure~\ref{fig:cosmo} shows the lensing measurements, $\Delta \Sigma(R)$, in the left panel, from DESY3+KiDS1000 (green) and from HSC (orange) and the clustering measurements, $w_{\rm p}(R)$, from BOSS are shown in the right panel (black). The data and measurement methods are detailed in Sections~\ref{sec:data} and~\ref{sec:measurements}. 
Measurements of $\Delta\Sigma$ account for lens magnification and systematic uncertainty due to photometric redshift and shear calibration. The best fits to the data are shown for the \textit{Planck} cosmology in red and the \textit{Lensing} cosmology in blue, derived similarly to Section~\ref{sec:joint} using the {\sc Dark Emulator}, now with an additional systematic parameter, $A$, which captures any overall inconsistency between the lensing and clustering measurements across all scales for each lens bin. The shaded region corresponds to the error on the model given by the $1\sigma$ error from the posterior distribution computed from the MCMC chains. 
The corresponding best fit values of $A$ are presented in Table~\ref{tab:Alens} and shown in Figure~\ref{fig:alens}. The fits are reasonable: $\chi^2=[33.3, 31.9, 32,3, 13,8]$ (\textit{Planck} cosmology) and $\chi^2=[31.0, 31.8, 29.5 12.9]$ (\textit{Lensing} cosmology) and assuming an effective DoF of 19-21. Overall there is a slightly better fit for the \textit{Lensing} cosmology, with a $\Delta\chi^2\sim5$ and $\Delta\chi^2\sim2$ when removing bin C1.  Assuming a low-$S8$ cosmology we find a smaller value of $A$ and therefore greater consistency between the lensing and clustering measurements within this model. Full discussion of these joint fits are detailed in Section~\ref{sec:Alens}.

\section*{Affiliations}
$^{1}$Kavli Institute for Cosmology, University of Cambridge, Madingley Road, Cambridge CB3 0HA, UK\\
$^{2}$Institute of Astronomy, University of Cambridge, Madingley Road, Cambridge, CB3 0HA\\
$^{3}$Kobayashi-Maskawa Institute for the Origin of Particles and the Universe (KMI), Nagoya University, Nagoya, 464-8602, Japan\\
$^{4}$Kavli Institute for the Physics and Mathematics of the Universe (WPI), The University of Tokyo, Chiba 277-8583, Japan \\
$^{5}$ Institute for Astronomy, University of Edinburgh, Royal Observatory, Blackford Hill, Edinburgh, EH9 3HJ, UK \\
$^{6}$Ruhr-Universit$\ddot{a}$t Bochum, Astronomisches Institut, German Centre for Cosmological Lensing (GCCL), Universitätsstr. 150,44801, Bochum, Germany\\
$^{7}$Department of Physics, University of California, Berkeley, CA 94720 \\
$^{8}$Physics Division, Lawrence Berkeley National Laboratory, Berkeley, CA \\
$^{9}$Kavli Institute for Particle Astrophysics and Cosmology, P. O. Box 2450, Stanford University, Stanford, CA 94305, USA \\
$^{10}$ Department of Physics, Stanford University, 382 Via Pueblo Mall, Stanford, CA 94305, USA\\
$^{11}$ SLAC National Accelerator Laboratory, Menlo Park, CA 94025, USA\\
$^{12}$ Max Planck Institute for Extraterrestrial Physics, Giessenbachstrasse, 85748 Garching, Germany\\
$^{13}$ Universit\"ats-Sternwarte, Fakult\"at f\"ur Physik, Ludwig-Maximilians Universit\"at M\"unchen, Scheinerstr. 1, 81679 M\"unchen, Germany\\
$^{14}$ The Inter-University Centre for Astronomy and Astrophysics, Post bag 4, Ganeshkhind, Pune 411007, India \\
$^{15}$ Faculty of Physics, Ludwig-Maximilians-Universit\"at, Scheinerstr. 1, 81679 Munich, Germany\\
$^{16}$ Center for Cosmology and Astro-Particle Physics, The Ohio State University, Columbus, OH 43210, USA\\
$^{17}$ Santa Cruz Institute for Particle Physics, Santa Cruz, CA 95064, USA\\
$^{18}$ Leiden Observatory, Leiden University, Niels Bohrweg 2, 2333 CA Leiden, the Netherlands\\
$^{19}$ Argonne National Laboratory, 9700 South Cass Avenue, Lemont, IL 60439, USA\\
$^{20}$ E.A Milne Centre, University of Hull, Cottingham Road, Hull, HU6 7RX, United Kingdom \\
$^{21}$ Department of Physics, Northeastern University, Boston, MA 02115, USA\\
$^{22}$ Laboratory of Astrophysics, \'Ecole Polytechnique F\'ed\'erale de Lausanne (EPFL), Observatoire de Sauverny, 1290 Versoix, Switzerland\\
$^{23}$ Department of Physics, Carnegie Mellon University, Pittsburgh, Pennsylvania 15312, USA\\
$^{24}$ Department of Physics, Duke University Durham, NC 27708, USA\\
$^{25}$ California Institute of Technology, 1200 East California Blvd, MC 249-17, Pasadena, CA 91125, USA\\
$^{26}$ Institut d'Estudis Espacials de Catalunya (IEEC), 08034 Barcelona, Spain\\
$^{27}$ Institute of Space Sciences (ICE, CSIC),  Campus UAB, Carrer de Can Magrans, s/n,  08193 Barcelona, Spain\\
$^{28}$ Fermi National Accelerator Laboratory, P. O. Box 500, Batavia, IL 60510, USA\\
$^{29}$ Department of Physics and Astronomy, University of Pennsylvania, Philadelphia, PA 19104, USA\\
$^{30}$ Department of Physics, The Ohio State University, Columbus, OH 43210, USA\\
$^{31}$ Jet Propulsion Laboratory, California Institute of Technology, 4800 Oak Grove Dr., Pasadena, CA 91109, USA\\
$^{32}$ Institut de F\'{\i}sica d'Altes Energies (IFAE), The Barcelona Institute of Science and Technology, Campus UAB, 08193 Bellaterra (Barcelona) Spain\\
$^{33}$ Center for Astrophysical Surveys, National Center for Supercomputing Applications, 1205 West Clark St., Urbana, IL 61801, USA\\
$^{34}$ Department of Astronomy, University of Illinois at Urbana-Champaign, 1002 W. Green Street, Urbana, IL 61801, USA\\
$^{35}$ Department of Astronomy, University of Geneva, ch. d'\'Ecogia 16, CH-1290 Versoix, Switzerland\\
$^{36}$ Department of Physics \& Astronomy, University College London, Gower Street, London, WC1E 6BT, UK\\
$^{37}$ Department of Applied Mathematics and Theoretical Physics, University of Cambridge, Cambridge CB3 0WA, UK\\
$^{38}$ Instituto de F\'isica Gleb Wataghin, Universidade Estadual de Campinas, 13083-859, Campinas, SP, Brazil\\
$^{39}$ Center for Gravitational Physics, Yukawa Institute for Theoretical Physics, Kyoto University, Kyoto 606-8502, Japan\\
$^{40}$ Department of Astronomy and Astrophysics, University of Chicago, Chicago, IL 60637, USA\\
$^{41}$ Kavli Institute for Cosmological Physics, University of Chicago, Chicago, IL 60637, USA\\
$^{42}$ Centro de Investigaciones Energ\'eticas, Medioambientales y Tecnol\'ogicas (CIEMAT), Madrid, Spain\\
$^{43}$ Brookhaven National Laboratory, Bldg 510, Upton, NY 11973, USA\\
$^{44}$ D\'{e}partement de Physique Th\'{e}orique and Center for Astroparticle Physics, Universit\'{e} de Gen\`{e}ve, 24 quai Ernest Ansermet, CH-1211 Geneva, Switzerland\\
$^{45}$ Laborat\'orio Interinstitucional de e-Astronomia - LIneA, Rua Gal. Jos\'e Cristino 77, Rio de Janeiro, RJ - 20921-400, Brazil\\
$^{46}$ Institute of Cosmology and Gravitation, University of Portsmouth, Portsmouth, PO1 3FX, UK\\
$^{47}$ Center for Theoretical Physics, Polish Academy of Sciences, al. Lotników 32/46, 02-668 Warsaw, Poland\\
$^{48}$ Instituto de Astrofisica de Canarias, E-38205 La Laguna, Tenerife, Spain\\
$^{49}$ Universidad de La Laguna, Dpto. AstrofÃ­sica, E-38206 La Laguna, Tenerife, Spain\\
$^{50}$ Physics Department, William Jewell College, Liberty, MO, 64068\\
$^{51}$ Astronomy Unit, Department of Physics, University of Trieste, via Tiepolo 11, I-34131 Trieste, Italy\\
$^{52}$ INAF-Osservatorio Astronomico di Trieste, via G. B. Tiepolo 11, I-34143 Trieste, Italy\\
$^{53}$ Institute for Fundamental Physics of the Universe, Via Beirut 2, 34014 Trieste, Italy\\
$^{54}$ Observat\'orio Nacional, Rua Gal. Jos\'e Cristino 77, Rio de Janeiro, RJ - 20921-400, Brazil\\
$^{55}$ Department of Physics, University of Michigan, Ann Arbor, MI 48109, USA\\
$^{56}$ Hamburger Sternwarte, Universit\"{a}t Hamburg, Gojenbergsweg 112, 21029 Hamburg, Germany\\
$^{57}$ Kapteyn Astronomical Institute, University of Groningen, PO Box 800, 9700 AV Groningen, The Netherlands \\
$^{58}$ Department of Physics, IIT Hyderabad, Kandi, Telangana 502285, India\\
$^{59}$ Institute of Theoretical Astrophysics, University of Oslo. P.O. Box 1029 Blindern, NO-0315 Oslo, Norway\\
$^{60}$ Department of Astronomy, University of Michigan, Ann Arbor, MI 48109, USA\\
$^{61}$ Instituto de Fisica Teorica UAM/CSIC, Universidad Autonoma de Madrid, 28049 Madrid, Spain\\
$^{62}$ School of Mathematics and Physics, University of Queensland,  Brisbane, QLD 4072, Australia\\
$^{63}$ Department of Astrophysical Sciences, Princeton University, Peyton Hall, Princeton, NJ 08544, USA\\
$^{64}$ Australian Astronomical Optics, Macquarie University, North Ryde, NSW 2113, Australia\\
$^{65}$ Lowell Observatory, 1400 Mars Hill Rd, Flagstaff, AZ 86001, USA\\
$^{66}$ Departamento de F\'isica Matem\'atica, Instituto de F\'isica, Universidade de S\~ao Paulo, CP 66318, S\~ao Paulo, SP, 05314-970, Brazil\\
$^{67}$ George P. and Cynthia Woods Mitchell Institute for Fundamental Physics and Astronomy, and Department of Physics and Astronomy, Texas A\&M University, College Station, TX 77843,  USA\\
$^{68}$ Instituci\'o Catalana de Recerca i Estudis Avan\c{c}ats, E-08010 Barcelona, Spain\\
$^{69}$ Physics Department, 2320 Chamberlin Hall, University of Wisconsin-Madison, 1150 University Avenue Madison, WI  53706-1390\\
$^{70}$ Perimeter Institute for Theoretical Physics, 31 Caroline St. North, Waterloo, ON N2L 2Y5, Canada\\
$^{71}$Shanghai Astronomical Observatory (SHAO), Nandan Road 80, Shanghai 200030, China \\
$^{72}$University of Chinese Academy of Sciences, Beijing 100049, China\\
$^{73}$ Computer Science and Mathematics Division, Oak Ridge National Laboratory, Oak Ridge, TN 37831\\

\label{lastpage}												

\end{document}